\documentclass[12pt,epsf]{article}
\usepackage{verbatim}

\usepackage{dsfont}
\usepackage{sidecap}
\sidecaptionvpos{figure}{t}
\usepackage{subfigure}

\usepackage[english]{babel}

\usepackage{amssymb,amsmath}
\usepackage{graphicx, xcolor, varwidth}
\usepackage{setspace}
\usepackage[permil]{overpic}
\usepackage{cite}

\numberwithin{equation}{section}

\newcommand{\be}{\begin{equation}}
\newcommand{\ee}{\end{equation}}
\newcommand{\bea}{\begin{eqnarray}}
\newcommand{\eea}{\end{eqnarray}}
\newcommand{\bear}{\begin{eqnarray}}
\newcommand{\eear}{\end{eqnarray}}  
\newcommand{\beas}{\begin{eqnarray*}}
\newcommand{\p}{\partial}
\newcommand{\eeas}{\end{eqnarray*}}
\newcommand{\ba}{\begin{array}}
\newcommand{\ea}{\end{array}}

\newcommand{\del}{\nabla}

\newcommand{\pd}[2][1]{\ifnum#1=1 \frac{\partial}{\partial {#2}} \else
  \frac{\partial^#1}{\partial {#2}^{#1}}\fi}
\newcommand{\dpd}[2][1]{\ifnum#1=1 \dfrac{\partial}{\partial {#2}} \else
  \frac{\partial^#1}{\partial {#2}^{#1}}\fi}
\newcommand{\td}[2][1]{\ifnum#1=1 \frac{d}{d{#2}} \else
  \frac{d^#1}{d{#2}^{#1}}\fi}

\newcommand{\nbox}{{\,\lower0.9pt\vbox{\hrule \hbox{\vrule height 0.2 cm \hskip 0.19 cm \vrule height 0.2 cm}\hrule}\,}}

\newcommand{\ie}{{\it i.e.,}\ }

\newcommand{\para}[1]{\paragraph{#1}\mbox{}\\}

\newcommand{\bh}{\bar{h}}
\newcommand{\bx}{\bar{x}}

\newcommand{\half}{\tfrac{1}{2}}

\newcommand{\bz}{\bar{z}}

\newcommand{\bw}{\bar{w}}

\newcommand{\brho}{\bar{\rho}}

\newcommand{\bphi}{\bar{\phi}}

\newcommand{\ep}{\varepsilon}
\newcommand{\bvep}{\bar{\varepsilon}}


\begin{document}
\begin{spacing}{1.3}
\begin{titlepage}

\begin{center}
{\Large \bf Causality Constraints in \\ \vspace{.3cm}  Conformal Field Theory}

\vspace*{6mm}

Thomas Hartman, Sachin Jain, and Sandipan Kundu
\vspace*{6mm}

\textit{Department of Physics, Cornell University, Ithaca, New York\\}

\vspace{6mm}

{\tt hartman@cornell.edu, sj339@cornell.edu, kundu@cornell.edu}

\vspace*{6mm}
\end{center}
\begin{abstract}

Causality places nontrivial constraints on QFT in Lorentzian signature, for example fixing the signs of certain terms in the low energy Lagrangian.  In $d$ dimensional conformal field theory, we show how such constraints are encoded in crossing symmetry of Euclidean correlators, and derive analogous constraints directly from the conformal bootstrap (analytically). The bootstrap setup is a Lorentzian four-point function corresponding to propagation through a shockwave.  Crossing symmetry fixes the signs of certain log terms that appear in the conformal block expansion, which constrains the interactions of low-lying operators.  As an application, we use the bootstrap to rederive the well known sign constraint on the $(\p \phi)^4$ coupling in effective field theory, from a dual CFT. We also find constraints on theories with higher spin conserved currents. Our analysis is restricted to scalar correlators, but we argue that similar methods should also impose nontrivial constraints on the interactions of spinning operators.

\end{abstract}

\end{titlepage}
\end{spacing}

\vskip 1cm

\setcounter{tocdepth}{2}
\tableofcontents

\begin{spacing}{1.3}
\newpage 

\section{Introduction}

Quantum field theories that appear to have causal propagation in vacuum can violate causality in nontrivial states.  Requiring the theory to be causal in every state constrains the allowed interactions. For example, in a massless scalar theory $\mathcal{L} = -(\p \phi)^2 + \mu (\p \phi)^4$, causality fixes the sign of the coupling $\mu \geq 0$ \cite{Adams:2006sv}. This not only constrains individual QFTs, but also plays an essential role in the proof of the $a$ theorem, and therefore constrains the relationship between different fixed points \cite{Komargodski:2011vj}. A second example comes from gravity, where causality constrains the coefficients of higher curvature corrections \cite{Brigante:2008gz,Hofman:2009ug,Camanho:2009vw,Camanho:2014apa}.

Causality constraints appear to rely inherently on Lorentzian signature, for obvious reasons.  The examples above can be restated in $S$-matrix terms as positivity of a scattering cross section, once again a condition that makes sense only in Lorentzian signature.  It is an open question how these constraints arise in the Euclidean theory.  In some broad sense, this was answered long ago by Schwinger, Wightman, and \mbox{others \cite{schwinger}:} every unitary, Lorentz invariant and causal theory in Minkowski space can be analytically continued to a Euclidean QFT satisfying reflection positivity, crossing symmetry, and Euclidean invariance. The converse holds as well, so good Euclidean theories are in one-to-one correspondence with good Lorentzian theories (assuming some bounds on the growth of correlators) \cite{OS}. However, this does not answer the question of what actually goes wrong in the Euclidean theory if, say, $(\p\phi)^4$ has the wrong sign, or we try to implement an RG flow between fixed points that violate the $a$ theorem.

In this paper, we answer a version of this question in conformal field theory in $d>2$ spacetime dimensions, using methods from the conformal bootstrap. The bootstrap has recently led to significant progress in deriving general constraints on the space of CFTs, from two general directions. First, in Euclidean signature, crossing has been implemented by computer to derive exclusion bounds on the allowed spectrum of low-lying operators (for example \cite{Rattazzi:2008pe, ElShowk:2012ht,Beem:2013qxa} and many others).  Second, in Lorentzian signature, the crossing equation as one operator approaches the lightcone of another operator is solvable, and fixes the dimensions of certain high spin operators in terms of the low-lying contributions to the crossing equation \cite{Komargodski:2012ek,Fitzpatrick:2012yx} (see also \cite{Fitzpatrick:2014vua,Alday:2014tsa,Vos:2014pqa,Fitzpatrick:2015qma,Alday:2015ota,Kaviraj:2015xsa,Kaviraj:2015cxa,Fitzpatrick:2015zha}).  This approach is Lorentzian, but operators are spacelike separated, or nearly so.  A natural guess is that causality constraints are somehow related to crossing symmetry when operators are timelike separated.

We will formulate crossing at timelike separation and confirm this guess. We study the scalar four-point function,
\be
\langle \psi OO \psi \rangle \ ,
\ee
which can be viewed as a two-point function of $O$ in the background produced by $\psi$. If all four operators are spacelike separated, then causality in the form $[O,O]=0$ is obviously related to crossing symmetry, which exchanges $O \leftrightarrow O$.  But when $O$ and $\psi$ are timelike separated, causality becomes a subtle question of where singularities appear in this position-space correlator.  We show that reflection positivity and crossing symmetry guarantee causality, \ie they prevent the lightcone from shifting acausally in the $\psi$ background.  Furthermore, crossing fixes the sign of certain finite corrections just before the lightcone.  This constrains the interactions of light operators in a nontrivial way, and is analogous to the causality constraints mentioned above. In CFT the constraints take the form (ignoring normalizations)
\be
\sum_i c_{OOX_i} c_{\psi \psi X_i } > 0 \ ,
\ee
where $O$ and $\psi$ are scalar operators, the sum is over minimal-twist operators $X_i$ with spin $>1$,  and  the $c_{XYZ}$ are three-point couplings.  We also give a formula for the positive quantity on the left-hand side, in terms of  
OPE data in the dual channel, \ie the couplings $|c_{O\psi P}|^2$.
It is an integral of a manifestly-positive commutator over the Regge regime of the correlator.
This can be viewed as a sum rule that relates the lightcone limit of the 4-point function to the Regge limit.  It is reminiscent of the optical theorem (which appears in previous work on both causality \cite{Adams:2006sv} and the lightcone bootstrap \cite{Komargodski:2012ek}) but we work entirely in position space.

Our approach is quite different from the old reconstruction theorems relating good Lorentzian theories to good Euclidean theories.  We constrain the interactions of light operators, independent of the rest of the theory.  This is similar to the point of view in effective field theory that UV consistency imposes IR constraints. 

A version of our position-space optical theorem holds also for the correlation functions of non-conformal quantum field theories. This relates the lightcone limit to a Regge-like limit in any relativistic QFT, at least in principle, but it remains to be seen whether any of the contributions can actually be calculated without conformal symmetry. It would be very interesting to find other useful examples.

\subsection{Holographic motivation and $(\p \phi)^4$}
We consider only external scalars in this paper, and we do not assume large $N$ in the derivation of positivity or the sum rule.  However we are partly motivated by the possibility that similar methods, applied to external operators with spin, may  shed some light on emergent geometry in large-$N$ CFTs. 

In any theory of quantum gravity, the low energy behavior of gravitons is governed by the Einstein-Hilbert action plus terms suppressed by a dimensionful scale $M_{2}$. Schematically,
\be\label{eha}
S  \sim \frac{1}{16\pi G_N}\int \sqrt{-g}\left(-2\Lambda+R + \frac{c_2}{(M_{2})^2}R^2 + \cdots\right) \ ,
\ee
where $c_2$ is an order 1 constant. Suppose $M_{2} \ll M_{Planck}$, and that the particular choice of $R^2$ term is ghost-free (for example the Gauss-Bonnet term in five dimensions). Standard effective field theory suggests that there is new physics at the scale $M_{2}$, but does not actually require it --- without taking causality constraints into account, this Lagrangian alone is perturbatively consistent up to a scale parametrically higher than $M_{2}$. However, it violates causality in nontrivial backgrounds. This was first shown for $c_2$ outside a certain range \cite{Brigante:2008gz,Hofman:2009ug}, and more recently extended to any non-zero value of $c_2$ by Camanho, Edelstein, Maldacena, and Zhiboedov (CEMZ) \cite{Camanho:2014apa}.  The implication of the CEMZ paper is not that $c_2$ actually vanishes, but that there must be new physics at the scale $M_2$, well below the naive breakdown of \eqref{eha}.

The holographic dual of the statement that any theory of gravity is governed by \eqref{eha} at low energies is that any large-$N$ CFT should have a very specific set of stress tensor correlation functions:
\be\label{ttcon}
\langle T_{\mu\nu}T_{\rho\sigma} \cdots \rangle_{CFT} = \langle T_{\mu\nu}T_{\rho\sigma} \cdots \rangle_{Einstein} + \cdots
\ee
The first term on the right-hand side is the correlator computed from the Einstein action in the bulk, say by Witten diagrams, and the subleading terms are suppressed by the dimension of `new physics' operators. Clearly the statement only makes sense if the spectrum of low-dimension single trace operators is sparse, so that the subleading terms are suppressed.

That some statement along these lines should hold in CFT has been suggested since the early days of AdS/CFT. The authors of \cite{Heemskerk:2009pn} stated a detailed conjecture, and took a major step towards deriving it from CFT by setting up and partially proving a scalar version of the conjecture, using the conformal bootstrap. Any argument for the stress tensor version \eqref{ttcon} is still lacking (but see \cite{Beem:2013qxa,Alday:2014tsa} for suggestive results from the supersymmetric bootstrap). The story for scalars, initiated in \cite{Heemskerk:2009pn} and developed for example in \cite{Heemskerk:2010ty,Fitzpatrick:2010zm,Penedones:2010ue,ElShowk:2011ag,Fitzpatrick:2011ia,Fitzpatrick:2011hu,Fitzpatrick:2011dm,Fitzpatrick:2012cg,Goncalves:2014rfa,Alday:2014tsa,Hijano:2015zsa}, is that effective field theories in the bulk are in one-to-one correspondence with solutions of crossing symmetry in CFT, order by order in $1/N$.  For each interaction that can be added to the scalar action in the bulk, 
\be
S = -\int d^Dx (\p \phi)^2 + \cdots \ ,
\ee
there is a corresponding perturbative solution to crossing symmetry.   Equivalently, for each flat-space scalar $S$-matrix, there is a corresponding crossing-symmetric CFT correlator in Mellin space \cite{Mack:2009mi,Mack:2009gy,Penedones:2010ue,Fitzpatrick:2011ia}.

Causality constraints, however, were missing from the CFT side of this picture.  In the bulk, causality (or analyticity) dictates that certain interactions come with a fixed sign, most notably $(\p\phi)^4$ in a scalar theory with a shift symmetry \cite{Adams:2006sv}.  This can be translated, by the above technology, into a constraint on CFT data, but there was previously no direct CFT derivation. We will show that the $(\p \phi)^4$ interaction in the bulk introduces a log term in the dual CFT that is constrained by our arguments, reproducing \cite{Adams:2006sv}.

For scalars, causality constraints give inequalities.  But for gravity, causality constraints can be expected to play an even more central role. Brigante et al \cite{Brigante:2008gz} first used causality to derive inequalities for the coefficients of $R^2$ gravity.  Soon afterward, Hofman and Maldacena  showed that stress tensor three-point functions in CFT must obey the same inequalities, by requiring a positive energy flux \cite{Hofman:2008ar,Hofman:2009ug}. For example, in $d=4$ with $\mathcal{N}=1$ supersymmetry, the constraint is $|a-c|\leq c/2$ where the anomaly coefficients $a$ and $c$ are the two independent constants determining $\langle TTT\rangle$.  

The more recent CEMZ constraints \cite{Camanho:2014apa} are much stronger, implying $a \approx c$ up to corrections suppressed by the mass of higher spin particles. These new constraints have not been derived from CFT; to do so would be a particularly interesting subcase of \eqref{ttcon}. The gravity argument \cite{Camanho:2014apa} was based on causality in a nontrivial background, so it seems likely that causality constraints on four-point or higher-point functions should play a role in the CFT derivation of \eqref{ttcon}.  Unlike the Hofman-Maldacena constraints, the CEMZ constraints do not hold in every CFT. They require at least one additional assumption, that the low-lying spectrum is sparse, so that corrections to the Einstein correlators are suppressed.  This is an assumption that can plausibly be implemented in the conformal bootstrap, which is already organized as an expansion in scaling dimension (or twist). So an optimistic possibility is that methods similar to those developed here, extended to external spinning operators and combined with an assumption of large $N$ and a sparse spectrum, are a step towards a CFT derivation of $a \approx c$.  This is only speculation. We do not derive any actual bounds for external spinning operators, but we do discuss the structure of the expected constraints.

This philosophy parallels recent developments in 3d gravity/2d CFT, where the conjecture \eqref{ttcon} is trivial on the plane (due to Virasoro symmetry) but nontrivial at higher genus.  The simplest genus-one analogue is the Cardy formula: crossing symmetry (\ie modular invariance) of the genus one partition function leads to a CFT derivation of black hole entropy at high temperatures \cite{Strominger:1997eq}, and adding to this derivation the assumption of a sparse spectrum extends the match to all temperatures \cite{Hartman:2014oaa}. A similar approach has been applied to entanglement entropy \cite{Hartman:2013mia,Asplund:2014coa} and correlation functions of primary operators \cite{Fitzpatrick:2014vua,Asplund:2014coa,Fitzpatrick:2015zha} using conformal bootstrap methods very similar to those deployed in this paper. 

There is also a close connection to the Maldacena-Shenker-Stanford bound on chaos \cite{Maldacena:2015waa}, as applied to CFT. The chaos bound constrains data that, in general, cannot be accessed in the OPE, whereas our bound constrains low-dimension couplings that dominate the OPE in the lightcone limit.  We discuss this in detail below, and derive a new version of the chaos bound that applies in the perturbative, lightcone regime: the dominant operator exchange near the lightcone must have spin $\leq 2$. This implies, also, that a CFT cannot have a finite number of higher-spin conserved currents, in any spacetime dimension.  This was proved by Maldacena and Zhiboedov using other methods in $d=3$ \cite{Maldacena:2011jn} and extended to higher dimensions by an algebraic argument in \cite{Boulanger:2013zza}.

\subsection{Outline}

Section 2 is a very brief version of the main technical argument.  Section 3 is a pedagogical review of causality in position-space quantum field theory. Most or all of it was known in the 70's so it can be safely skipped by experts in this topic. In section 4, we review the conformal bootstrap in Euclidean signature and discuss its extension to Lorentzian signature with timelike separated points.  In section 5, we define a shockwave state in CFT and show that the 4-point function is causal.  The correlator is shown to be analytic in a certain regime of complex-$x_i$, which is used in section 6 to derive the sign constraints and sum rule.  Finally in section 7 (which is the only section where we assume large $N$) we derive the $(\p \phi)^4$ constraint \cite{Adams:2006sv} from a dual CFT.


\section{Brief argument for log bounds}\label{s:brief}

The main argument is simple, but the background and technical details are both lengthy, so we start with a concise argument for why log coefficients are constrained by the bootstrap. Details are mostly omitted, so this section is intended for experts who are already familiar with all of the ingredients and want a shortcut to the main result, or as a reference while reading the rest of the paper.

Consider the normalized four-point function
\be\label{gzzbb}
\overline{G}(z,\bz) =\frac{ \langle \psi(0) O(z,\bz) O(1) \psi(\infty)\rangle}{\langle O(z,\bz) O(1)\rangle} \ ,
\ee
where $z,\bz$ are conformal cross ratios.   For $\bz = z^*$, this is a Euclidean correlator, but more generally, $z$ and $\bz$ are independent complex numbers.  Causality is encoded in the analytic structure of $G(z,\bz)$ as a function on some multi-sheeted cover of $\mathbf{C} \times\mathbf{C}$ (section \ref{s:causality}). We will define a particular state, the shockwave state (section \ref{s:cshock}), and demonstrate the following:

\begin{enumerate}
\item   Let
\be
z = 1 + \sigma , \qquad \bz = 1 + \eta \sigma
\ee
where $\eta \ll 1$ is real and $\sigma$ is complex.  Then the CFT is causal in the shockwave state, in the sense that 
\be
\langle \Psi | [O,O] | \Psi \rangle
\ee
vanishes outside the lightcone, if and only if 
\be\label{wgbk}
\widehat{G}_\eta(\sigma) \equiv \overline{G}(e^{-2\pi i}(1+\sigma), 1 + \eta \sigma)
\ee
is an analytic function of $\sigma$ for Im $\sigma \geq 0$ in a neighborhood of $\sigma = 0$. The $e^{-2\pi i}$ here is responsible for ordering operators correctly as in \eqref{gzzbb}. The commutator is computed by the discontinuity across a cut, so it vanishes where this function is analytic. The limit $\eta \to 0$ (fixed $\sigma$) is the lightcone limit, and the limit $\sigma \to 0$ (fixed $\eta$) is the Regge limit.\footnote{Throughout the paper, the term Regge limit refers somewhat loosely to any regime that has $z,\bz \sim 1$ and is outside the lightcone limit.}

\item If the CFT is reflection-positive, then there is an OPE channel which converges in the region just described. It follows that the correlator is analytic on this region, and therefore causal (section \ref{s:cshock}).

\item In the lightcone limit $\eta \ll |\sigma|$, we can expand in the $OO$ OPE, with the dominant contributions from low-twist operators.  Suppose the minimal-twist operator is the stress tensor, and set $d=4$ (more general cases are considered below). Then
\be
G(z,\bz) = 1 + c_{OOT}c_{\psi\psi T} (1-\bz) \tilde{g}_{4,2}(1-z) + O((1-\bz)^2) \ ,
\ee
where $\tilde{g}_{4,2}$ is the lightcone conformal block for stress-tensor exchange, known in any spacetime dimension (equation \eqref{tlblock}), and the $c$'s are OPE coefficients. Plugging in the kinematics \eqref{wgbk} and expanding for $\eta \ll |\sigma|\ll 1$, 
\be\label{tche}
\widehat{G}_\eta(\sigma) = 1 - i \lambda \frac{\eta}{\sigma} + O(\eta^2) \ ,
\ee
where $\lambda \propto c_{OOT}c_{\psi\psi T}$. The usual OPE has only positive powers of $\eta,\sigma$, but here we see $\sigma^{-1}$ --- this negative power is the key observation that leads to constraints.  It comes from a term $\sim \log\left( \sigma e^{-2\pi i}\right) $ in the lightcone block for the shockwave kinematics. More generally the power for spin-$\ell$ exchange is $\sigma^{-\ell+1}$, so conformal blocks for operators of spin $>1$ are greatly enhanced on the second sheet.

\item Now we combine points \textit{(2)} and \textit{(3)} above to derive the sum rule for the log coefficient $\lambda$ (section \ref{s:logs}).  Analyticity of the position-space correlator implies
\be\label{introint}
\oint d\sigma ( \widehat{G}_\eta(\sigma) -1 ) = 0
\ee
along a closed path.  Choose the path to be a semicircle of radius $R$, just above the origin of the $\sigma$ plane:
\be\label{introdpath}
\includegraphics{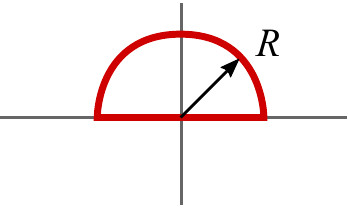}
\ee
We fix $\eta \ll R \ll 1$, so $\sigma$ near the edge of the semicircle is the lightcone limit ($\bz \to 1$ with $z$ fixed), while $\sigma$ near the origin is the Regge limit ($\bz,z \to 1$). (A contour integral like \eqref{introint} relates the lightcone limit to a Regge-like limit in any relativistic quantum field theory, not only CFTs, and may be useful in other contexts.)

In a CFT, the contribution from the semicircle can be computed from \eqref{tche}; taking the real part extracts the residue. Then \eqref{introint} becomes
\be\label{ibsumrule}
\pi \lambda \eta = \int_{-R}^R dx \, \mbox{Re}\,  (1 - \widehat{G}_\eta(x)) \ .
\ee
This integrand can also be written as the real part of a commutator.

\item Finally, expand the correlator in the $s$ channel $O(z,\bz) \to \psi(0)$,
\be
G(z,\bz) \propto \sum_{\Delta,\ell}a(\Delta,\ell) \left[ z^{\half(\Delta-\ell)} \bz^{\half(\Delta+\ell)} + (z \leftrightarrow \bz)\right] \ .
\ee
In a reflection-positive CFT, all of the coefficients in this expansion are positive: $a(\Delta,\ell) > 0$ (section \ref{s:lope}).  Thus sending $z \to e^{-2\pi i }z$ can only \textit{decrease} the magnitude of the correlator. Restoring all the correct prefactors this implies
\be\label{regbn}
\mbox{Re} \, \widehat{G}_\eta(x)\le \mbox{Re} \,  \bar{G}(1+x, 1+\eta x) \lesssim 1
\ee
for real  $|x| \ll 1$. The $s$ channel argument only applies for $x<0$, but a similar result in the $u$ channel $O(z,\bz) \to \psi(\infty)$ gives the same inequality for $x>0$. Corrections on the right-hand side of \eqref{regbn} are suppressed by positive powers of $\eta$ and $\sigma$, so they do not affect the sum rule.  Therefore the integrand in \eqref{ibsumrule} is positive.

\noindent This proves $\lambda>0$.  Actually, for stress tensor exchange, this constraint is trivial; the coefficient is fixed by the conformal Ward identity to be
\be
\lambda_T \sim \frac{\Delta_O\Delta_\psi}{c} \ ,
\ee
which is obviously positive.  But in slightly different situations discussed below the constraints obtained by identical reasoning are nontrivial.

\item The constraint can also be stated using the maximum modulus principle: the magnitude of an analytic function in a region $D$ is bounded by the maximum magnitude on $\p D$.  There is simply no way for a function of the form \eqref{tche} to be analytic inside the semicircle in \eqref{introdpath}, bounded by 1 on the real line, and have $\lambda <0$.  This can be easily checked by finding the maximum of this function along the semicircle with $|\sigma|=R$, and comparing to the maximum along the semicircle $|\sigma| = R-\delta R$, which must be smaller. Similar reasoning rules out the possibility that the dominant exchange has $\ell >2$, because in this case both choices of sign violate the maximum modulus principle. 
This version of the argument is inspired by the `signalling' argument in \cite{Camanho:2014apa} and the chaos bound \cite{Maldacena:2015waa}.

\end{enumerate}

Note that we do not assume causality; we use the conformal block expansion and reflection positivity to derive causality, then apply this result to derive the log bound.  If we had simply assumed causality then the argument for log bounds would be significantly shorter.

We go through all of these steps in great detail in the rest of the paper.


\section{Causality review}\label{s:causality}

Causality requires commutators to vanish outside the light-cone:\footnote{
Here is the standard argument for \eqref{comvan}: The theory cannot be quantized in a way consistent with boost invariance if \eqref{comvan} is violated.  To see this, add a local perturbation to the Hamiltonian, $H_{int} = \lambda O_2(x,t) \delta(x)\delta(t)$, and calculate in the interaction picture
\begin{align}
\langle \Omega| O_1(x) | \Omega\rangle = \langle 0 | e^{i \int_{-\infty}^t H_{int}dt} &O_1(x) e^{-i \int_{-\infty}^t H_{int}dt}|0\rangle \\
&= \langle 0|O_1(x)|0\rangle + \lambda \Theta(t) \langle 0 |[O_1(x), O_2(0)]|0\rangle + \cdots  \ .\notag
\end{align}
For spacelike separation, the step function $\Theta(t)$ is not invariant under boosts, so different coordinate systems disagree about the $O(\lambda)$ term if it is non-zero. The same argument can be repeated in any state, so \eqref{comvan} holds as an operator equation.}
\be\label{comvan}
[O_1(x), O_2(0)] = 0 , \qquad x^2 > 0 , \qquad x \in R^{d-1,1} \ ,
\ee
where $O_{1,2}$ are local operators inserted in Minkowski space. In this section we will review how this requirement is encoded in the analytic structure of correlation functions, first in a general Lorentz-invariant QFT and then in CFT.  This is an informal derivation of the position-space $i\epsilon$ prescription stated in standard references, for example the textbook by Haag \cite{haag}.

\subsection{Euclidean and Lorentzian correlators}

In a general Lorentz-invariant QFT, consider the Euclidean correlator on a plane,
\be
G(x_1, \dots, x_n) = \langle O_1(x_1) \dots O_n(x_n)\rangle \ .
\ee
This is a single valued, permutation invariant function of the positions
\be
x_i = (\tau_i, x_i^1, \dots, x_i^{d-1})  \in R^d \ .
\ee
Here $\tau$ is a direction, chosen arbitrarily, that will play the role of imaginary time.
$G$ is analytic away from coincident points, and has no branch cuts as long as all $n$ points remain Euclidean.  This reflects the fact that in Euclidean signature, operators commute:
\be
[O_1(x_1), \ O_2(x_2)] = 0 , \qquad x_{1,2}\in R^d \ , \qquad x_1 \neq x_2 \ .
\ee
Lorentzian correlators can be computed (or defined) by analytically continuing $\tau_i \to i t_i$, with $t_i$ real.  As functions of the complex $\tau_i$, the correlator has an intricate structure of singularities and branch cuts, leading to ambiguities in the analytic continuation. Each choice that we make in the analytic continuation translates into a choice of operator ordering in Lorentzian signature, so these ambiguities are responsible for non-vanishing commutators.  All of the Lorentzian correlators are analytic continuations of each other.\footnote{This is simple to prove: the Lorentzian correlators with various orderings are equal when points are spacelike separated, so it is a standard fact of complex analysis (the edge-of-the-wedge theorem) that they must all be related by analytic continuation in the positions.}
  
For instance, suppose we aim to compute the Lorentzian correlation function with all $t_i$'s zero except $t_2$, and displacement only in the direction $x^1\equiv y$, pictured in Lorentzian signature as follows:
\be\label{f:fourpoints}
\begin{gathered}\includegraphics{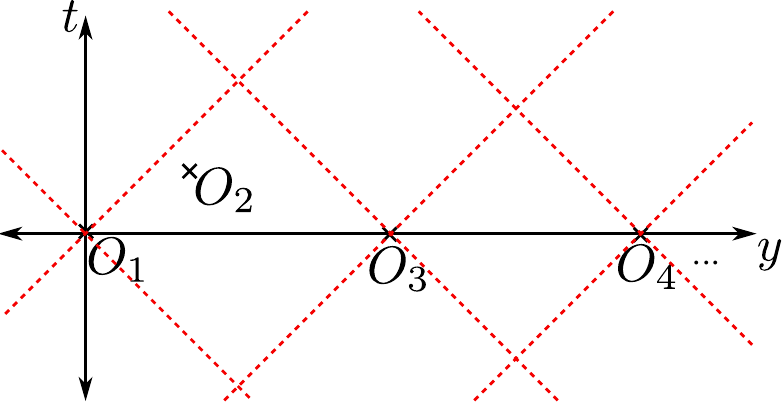}\end{gathered}
\ee
The correlator, viewed as a function of complex $\tau_2$ with all other arguments held fixed, has singularities along the imaginary-$\tau_2$ axis where $O_2$ hits the light cones of the other operators:
\be\label{f:t1bc}
\begin{gathered}\includegraphics{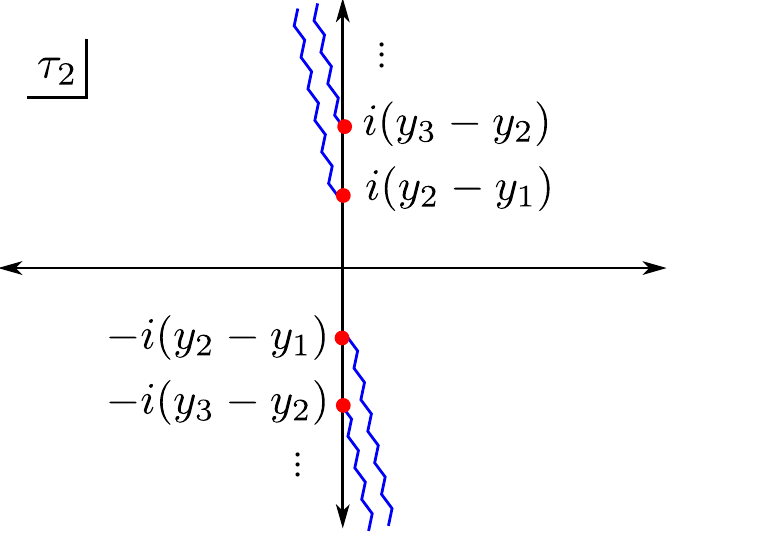}\end{gathered}
\ee
In an interacting theory, these singularities (red dots) are branch points, and we will orient the branch cuts (blue) so that they are `almost vertical' as in the figure.  In order to compute the correlator when $O_2$ is timelike separated from other operators, we need to continue from the point $\tau_2=t_2$ on the positive real axis to the point $\tau_2=it_2$ on the imaginary axis, which is above some light-cone singularities.  Each time we pass a singularity, we must choose whether to pass to the right or to the left.  Assume without loss of generality that $t_2 > 0$.  Then passing to the right of a singularity puts the operators into time ordering in the resulting Lorentzian correlator, and passing to the left puts the operators in anti-time-order.  

For example, suppose $O_2$ is in the future light cones of $O_1$ and $O_3$, but is spacelike separated from other operators. Then, starting from the single-valued Euclidean correlator, we can choose to go to Lorentzian signature along four different contours:
\be\label{f:t1c}
\begin{gathered}\includegraphics{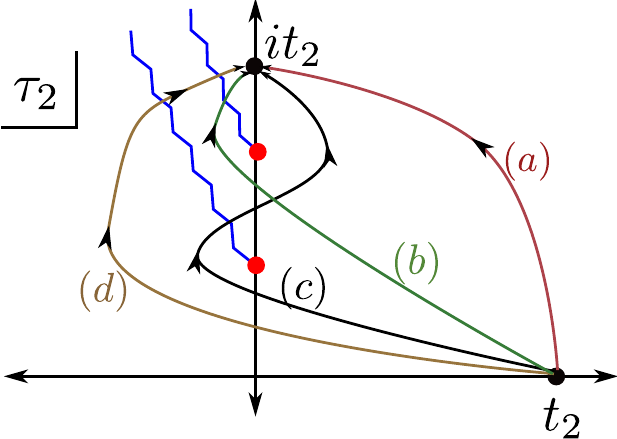}\end{gathered}
\ee
By choosing a contour, we mean that the analytic continuation is done in a way that is continuous along the given contour. These correspond, respectively, to the Lorentzian correlators
\bea
(a) & & \langle O_2 O_1 O_3 \cdots \rangle = \langle T[O_1 O_2 O_3 \cdots]\rangle  \notag \\
(b) & & \langle O_3 O_2 O_1 \cdots \rangle \notag\\
(c) & & \langle O_1 O_2 O_3 \cdots \rangle \notag\\
(d) & & \langle O_1 O_3 O_2 \cdots \rangle \ . \notag
\eea
$(a)$ is fully time-ordered, $(d)$ is fully anti-time-ordered, and the other two are mixed. If more than two operators were timelike-separated, then we would also need to worry about the ordering of the various branch cuts with respect to each other.

This recipe is motivated by the following observation. The branch cuts appear when operators become timelike separated, so to get a reasonable Lorentzian theory, the commutator must be equal to the discontinuity across the cut.  This implies that, for example, the function defined along contour $(a)$ in \eqref{f:t1c} differs from the function defined along $(b)$ by adding a commutator, $[O_2, O_3]$.  Combined with the fact that all Lorentzian correlators must continue to the (same) Euclidean correlator when operators are spacelike separated, this essentially fixes the prescription to what we have just described. See \cite{haag} and below for references to a full derivation.

\subsection{Causality}\label{ss:cc}

In order to diagnose whether a theory is causal, the actual value of the commutator is not needed -- the only question is where it is non-zero.  The answer, in the language of the analytically continued correlation functions, is that the commutator becomes non-zero when we encounter a singularity in the complex time plane and are forced to chose a contour.

The Euclidean correlator is singular only at coincident points.  This immediately leads to a causal Lorentzian correlator on the first sheet of the $\tau_2$ plane, which is the sheet pictured in \eqref{f:t1bc}.   To see this, note that a singularity at $x^2=0$ in Euclidean continues to a singularity at $x^2=0$ in Lorentzian, which is obviously on the light cone.   Thus, in the configuration discussed above, $\langle [O_2, O_1] O_3 \cdots\rangle$ and $\langle [O_2, O_3]O_1\cdots\rangle$ are manifestly causal: they become non-zero at the branch points drawn in that figure, which start precisely at the light cones. However, as we pass onto another sheet by crossing a branch cut, singularities could move.  For example, the commutator
\be
\langle O_1 [O_2, O_3] \cdots \rangle
\ee
becomes non-zero when we encounter the $O_3$ singularity along this contour:
\be\label{f:causalitycontour}
\begin{gathered}\includegraphics{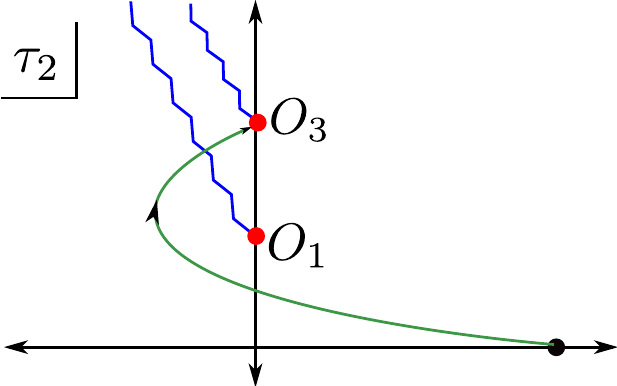}\end{gathered} \ .
\ee
It is not at all obvious that this singularity is at the $O_3$ lightcone, $\tau_2 = i (y_3-y_2)$. If, as we pass through the $O_1$ branch cut, this $O_2 \to O_3$ singularity shifts upwards along the imaginary axis, then the theory exhibits a time delay.  If it shifts downwards, then the commutator becomes non-zero earlier than expected (as a function of $t_2$) and the theory is acausal.

To summarize: Starting from a Euclidean correlator, causality on the first sheet is obvious.  The non-trivial statement about causality is a constraint on how singularities in the complex-$\tau$ plane move around as we pass through other light-cone branch cuts.

\subsection{Reconstruction theorems and the $i\epsilon$ prescription}

The Osterwalder-Schrader reconstruction theorem \cite{OS} states that well behaved Euclidean correlators, upon analytic continuation, result in Lorentzian correlators that obey the Wightman axioms. The definition of a well behaved Euclidean correlator is (i) analytic away from coincident points, (ii) $SO(d)$ invariant, (iii) permutation invariant, (iv) reflection positive, and (v) obeying certain growth conditions. Reflection positivity is the statement that certain correlators are positive and is discussed more below.

L\"{u}scher and Mack extended this result to conformal field theory defined on the Euclidean plane, showing that the resulting theory is well defined and conformally invariant not only in Minkowski space but on the Lorentzian cylinder \cite{Luscher:1974ez}.

A byproduct of these reconstruction theorems is a simple $i\epsilon$ prescription to compute Lorentzian correlators, with any ordering, from the analytically continued Euclidean correlators:
\be\label{iep}
\langle O_1(t_1, \vec{x}_1) O_2(t_2, \vec{x}_2) \cdots O_n(t_n, \vec{x}_n)\rangle = \lim_{\epsilon_j \to 0} \langle O_1(t_1-i\epsilon_1,\vec{x}_1) \cdots O_n(t_n-i\epsilon_n, \vec{x}_n)\rangle
\ee
where the limit is taken with $\epsilon_1>\epsilon_2>\cdots >\epsilon_n > 0$. The correlator on the rhs is analytic for any finite $\epsilon_k$ obeying these inequalities, which also confirms that the singularities we have been discussing always lie on the imaginary axis.

This $i\epsilon$ prescription is identical to our discussion above. It shifts the branch cuts to the left or right of the imaginary axis, and this enforces the contour choices that we described. For example contour $(c)$ in \eqref{f:t1c} corresponds to the $i\epsilon$ prescription
\be\label{f:upcut}
\begin{gathered}\includegraphics{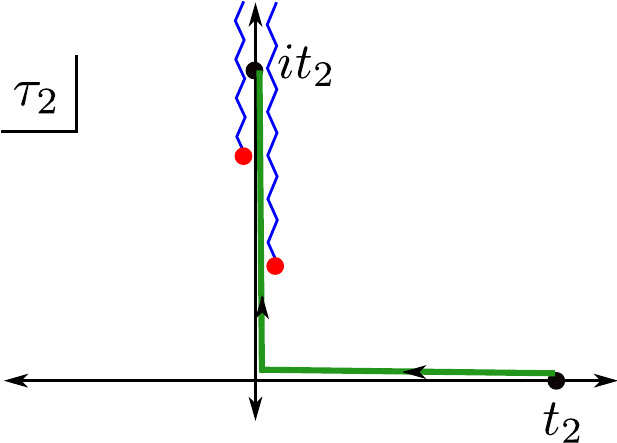}\end{gathered} \ ,
\ee
where the $i\epsilon$'s move the lightcone singularities off the imaginary-$\tau_2$ axis as indicated.\footnote{Note that inserting $i\epsilon$'s into the correlator is meaningless unless we also specify the positions of all the branch cuts.  On the complex $\tau_2$ plane, this does not lead to any confusion because the choice is always implicitly `straight upwards' as in the figure.  However when we write our correlators in terms of conformal cross ratios it is not obvious where to place the branch cuts on the $z,\bz$ planes. For this reason we will always give the contour description and avoid $i\epsilon$'s entirely in our calculations.}

In principle, the reconstruction theorem completely answers the question of when Euclidean correlators define a causal theory.  Our point of view, however, will be that we assume only some limited information about the CFT --- for example, there is some light operator of a particular dimension and spin, exchanged in a four-point function, perhaps with some particular OPE coefficients --- and we want to know whether this is compatible with causality.  This limited data may or may not come from a full QFT obeying the Euclidean axioms.  The reconstruction theorem does not answer this type of question in any obvious way. In other words, the reconstruction theorem tells us that causality violation in Lorentzian signature must imply some problem in Euclidean signature, but we want to track down exactly what that problem is.

\subsection{Examples}

\subsubsection*{Conformal 2-point function}

The Euclidean 2-point function in CFT is $(\tau^2 +x^2)^{-\Delta}$.  This is single-valued in Euclidean space, since the term in parenthesis is non-negative.  Using the $i\epsilon$ prescription, the Lorentzian correlators for $t_1>x_1$ are
\be\label{extoo}
\langle O(t_1, x_1)O(0,0)\rangle = \exp\left(-\Delta \log(-(t_1-i\epsilon)^2 + x_1^2)\right) =  e^{-i \pi  \Delta} (t_1^2-x_1^2)^{-\Delta} \ ,
\ee
and
\be\label{extoob}
\langle O(0,0) O(t_1, x_1) \rangle = \exp\left(-\Delta \log(-(t_1+i\epsilon)^2 + x_1^2)\right) =  e^{i \pi  \Delta} (t_1^2-x_1^2)^{-\Delta} \ ,
\ee
where we placed the branch cut of log on the negative real axis, as in \eqref{f:upcut}.

Alternatively, in the language of paths instead of $i\epsilon$'s, we start from the Euclidean correlator $(z\bz)^{-\Delta}$, where $z = x+i\tau$, $\bz = x - i \tau$.  To find the time-ordered Lorentzian correlator we set $\tau = t_2 e^{i \phi}$ and follow the path $\phi \in [0, \pi/2]$. The result agrees with \eqref{extoo}. The anti-time-ordering path goes the other way around the singularity at $z=0$, so it differs by $z \to z e^{-2\pi i}$, giving \eqref{extoob}.

\subsubsection*{Free 2-point function}
A free massless scalar in $d$ dimensions has $\Delta = d/2 - 1$. In even dimensions, this is an integer, so there are no branch cuts in the Lorentzian 2-point function.  It follows that $\langle [\phi(x), \phi(y)]\rangle = 0$ at timelike separation. Standard free field methods confirm that the commutator in even dimensions is supported only on the lightcone, $(x-y)^2=0$.

\subsection{CFT 4-point functions}\label{ss:cftcausality}
We now specialize to 4-point functions in a conformal field theory. Take the operators $O_1, O_3$ and $O_4$ to be fixed and spacelike separated at $\tau = 0$, while $O_2$ is inserted at an arbitrary time:
\be\label{xxpt}
x_1 = (0, \dots, 0) , \quad x_2 = (\tau_2, y_2, 0, \dots, 0), \quad x_3 = (0, 1, \dots, 0) , \quad x_4 = (0,\infty,0, \dots, 0) \ ,
\ee
with 
\be\label{xxptb}
0 < y_2 < \frac{1}{2} \ .
\ee 
This is similar to \eqref{f:fourpoints} but with $O_4$ moved to infinity.\footnote{$O(\infty) \equiv \lim_{y\to \infty}y^{2\Delta_O}O(y)$.} Only one of the operators is at $t \neq 0$, so the others are all spacelike separated. The conformal cross ratios are defined by
\be
u = \frac{x_{12}^2 x_{34}^2}{x_{13}^2 x_{24}^2} , \quad v = \frac{x_{14}^2 x_{23}^2}{x_{13}^2 x_{24}^2}  \ .
\ee
Another convenient notation is
\be\label{uvdef}
u = z\bz , \qquad v = (1-z)(1-\bz) \ ,
\ee
which for \eqref{xxpt} becomes\footnote{\eqref{uvdef} is invariant under $z \leftrightarrow \bz$, but we will always choose the solutions of the quadratic equation corresponding to \eqref{zzz}, so this distinguishes $z$ and $\bz$.}
\be\label{zzz}
z = y_2 + i \tau_2 , \qquad \bz = y_2 - i \tau_2 \ .
\ee
In Euclidean signature, $\tau_2$ is real and $\bz = z^*$. In Lorentzian signature, $z = y_2 - t_2$ and $\bz = y_2 + t_2$ are independent real numbers.

The Euclidean correlator $G(z,z^*)$ has the short-distance singularities
\be
G(z,z^*) \sim (z z^*)^{-\half(\Delta_1 + \Delta_2)} \quad \mbox{as} \quad z \to 0
\ee
and 
\be
G(z,z^*) \sim ((1-z)(1-z^*))^{-\half(\Delta_2 + \Delta_3)} \quad \mbox{as} \quad z \to 1 \ .
\ee
The various Lorentzian correlators are computed by analytic continuation $\tau_2 \to i t_2$. Denote by $G(z,\bz)$ the time-ordered correlator, defined by analytic continuation along the contour $(a)$ in \eqref{f:t1c}.  Then for real $z$ and $\bz$, and $O_2$ in the future lightcone of both $O_1$ and $O_3$, the contours in \eqref{f:t1c} correspond to the functions
\bea\label{confres}
&(a)& \qquad G(z,\bz) =  \langle O_2 O_1 O_3 O_4\rangle \\
&(b)& \qquad G(z,\bz)|_{(\bz-1)\to e^{-2\pi i}(\bz-1)} = \langle O_3 O_2 O_1 O_4 \rangle \notag \\
&(c)& \qquad G(z,\bz)|_{z \to e^{-2\pi i } z} = \langle O_1 O_2 O_3 O_4\rangle\notag \\
&(d)& \qquad G(z,\bz)|_{z \to e^{-2\pi i }z , (\bz - \bz_0) \to e^{-2\pi i}(\bz - \bz_0)} = \langle O_1 O_3 O_2 O_4\rangle \ .\notag
\eea
These follow from the fact that the first singularity above the real axis in \eqref{f:t1c} is $z=0$, and the second is $\bz = 1$. The subscripts indicate how to go around these singularities. In the last line, $\bz_0$ is defined to be the singularity of $G(ze^{-2\pi i }, \bz)$ as a function $\bz$:
\be
G(ze^{-2\pi i }, \bz) \to \infty \qquad \mbox{as} \qquad \bz \to \bz_0 \ ,
\ee  
coming from $O_3$, depicted in \eqref{f:causalitycontour}. According to the reconstruction theorems, it must lie on the real axis, $\mbox{Im}\, \bz_0$ = 0 (so that the singularity in the $\tau_2$ plane lies on the imaginary axis).  Comparing contours $(c)$ and $(d)$, the 4-point function is causal if and only if
\be
\mbox{Re}\, \bz_0 \geq 1 \ .
\ee

\section{The Lorentzian OPE}\label{s:lope}

In this section we review the Euclidean OPE in $d$-dimensional CFT, derive some consequences of reflection positivity, and discuss to what extent the OPE can be applied in Lorentzian correlators. For the simplest case where only one operator is timelike separated from the others, we show that there is a convergent OPE channel, and use it to link causality to reflection positivity.

\subsection{Conformal block expansion}
The operator product expansion in CFT is
\be
O_1(x_1) O_2(x_2) = \sum_k f_{12 k}(x_1-x_2) O_k(x_2) \ ,
\ee
where the function $f_{12k}$ is fixed by the three-point functions of primary operators together with the conformal algebra.  Applied inside a 4-point correlation function, the OPE gives the conformal block expansion:
\begin{align}\label{genblocks}
\langle O_1(x_1)O_2(x_2)O_3(x_3)&O_4(x_4)\rangle = \\
&  \frac{1}{x_{12}^{\Delta_1+\Delta_2}x_{34}^{\Delta_3+\Delta_4}}\left( x_{24}\over x_{14}\right)^{\Delta_{12}}\left(x_{14}\over x_{13}\right)^{\Delta_{34}}\sum_p c_{12p} c_{34p}g_{\Delta_p, \ell_p}^{\Delta_{12}, \Delta_{34}}(z, \bz)\notag \ , 
\end{align}
where $\Delta_{ij} = \Delta_i - \Delta_j$, $x_{ij} = x_i - x_j$, $c_{ijk}$ is the OPE coefficient. The sum is over conformal primaries, and the conformal block $g$ accounts for descendant contributions. The ordering of operators in the OPE coefficient $c_{ijk}$ is important, as it is antisymmetric for odd spin exchange:
\be
c_{ijk} = (-1)^{\ell_k}c_{jik} \ .
\ee
In $d=4$ \cite{Dolan:2000ut} the conformal blocks are known in terms of hypergeometric functions, and reproduced in appendix \ref{app:doblocks}.  In odd dimensions, the blocks are not known in closed form, but they can be computed efficiently by recursion relations \cite{ElShowk:2012ht}. The full conformal blocks are not needed for any of the results in this paper; we will only use the explicit form of the lightcone blocks, discussed below, which are known in all $d$.

\subsection{The Euclidean $z$-expansion}
Consider a 4-point function with two species of operators:
\be\label{gzc}
G(z,\bz) = \langle \psi(0) O(z,\bz) O(1) \psi(\infty)\rangle \ ,
\ee
where
\be
z = y_2 + i \tau_2 , \qquad \bz = y_2 - i \tau_2 \ ,
\ee
with
\be
0 < y_2 < \half \ .
\ee
In this configuration, the cross ratio \eqref{uvdef} is simply $(z, \bz) = x_2$.  

The OPE $O(z,\bz) \to \psi(0)$ leads to the $s$-channel conformal block expansion as in \eqref{genblocks},
\be\label{sblocks}
\mbox{$s$ channel:} \qquad G(z,\bz) = (z\bz)^{-\half(\Delta_O+\Delta_\psi)} \sum_p c_{\psi O p}c_{O\psi p}\, g_{\Delta_p, \ell_p}^{\Delta_{\psi O}, -\Delta_{\psi O}}(z,\bz) \ .
\ee
The sum converges for Euclidean points $\bz = z^*$ with $|z|<1$ \cite{Mack:1976pa,Pappadopulo:2012jk}. To see this, we can choose a sphere of radius $R$ with $|z|<R<1$, define states on the sphere by radial quantization, and reinterpret the conformal block expansion as a partial wave expansion, which must converge.

The $t$ channel $O(z,\bz) \to O(1)$, which can be obtained by first relabeling $x_1 \leftrightarrow x_3$ and then using \eqref{genblocks}, leads to the expansion
\be\label{tblocks}
\mbox{$t$ channel:} \qquad G(z,\bz) = ((1-z)(1-\bz))^{-\Delta_O}\sum_p c_{OOp} c_{\psi\psi p} \, g_{\Delta_p, \ell_p}^{0,0}(1-z,1-\bz) \ .
\ee
This converges for Euclidean points with $|1-z|<1$. Finally, expanding in the $u$ channel $O(z,\bz) \to \psi(\infty)$ gives 
\be\label{ublocks}
\mbox{$u$ channel:} \qquad G(z,\bz) = (z\bz)^{\half(\Delta_\psi - \Delta_O)} \sum_p c_{O\psi p}c_{\psi O p}\,  g_{\Delta_p, \ell_p}^{\Delta_{\psi O},- \Delta_{\psi O}}(\frac{1}{z}, \frac{1}{\bz}) \ ,
\ee
which is convergent for Euclidean $|z|>1$.

The operators $\psi$ and $O$ are real scalars, but everything in this section (and much of the rest of the paper) can be repeated for complex scalars and the correlator $\langle \psi(0)O(z,\bz)O^\dagger(1)\psi^\dagger(\infty)\rangle$. For complex operators, the ordering of OPE coefficients in the $s$, $t$, and $u$ expansions respectively is: $c_{\psi O p}c_{O^\dagger \psi^\dagger p}$, $c_{O^\dagger O p}c_{\psi \psi^\dagger p}$, and $c_{\psi^\dagger O p}c_{O^\dagger \psi p}$.

Crossing symmetry equates \eqref{sblocks}, \eqref{tblocks}, and \eqref{ublocks}.  However, for a given $z$, only two of the three expansions converge.  Even worse,  the $s$ and $u$ channels have no overlapping range of convergence. This is overcome by the $\rho$ expansion.

\subsection{The Euclidean $\rho$-expansion}

\begin{figure}
\centering
\includegraphics{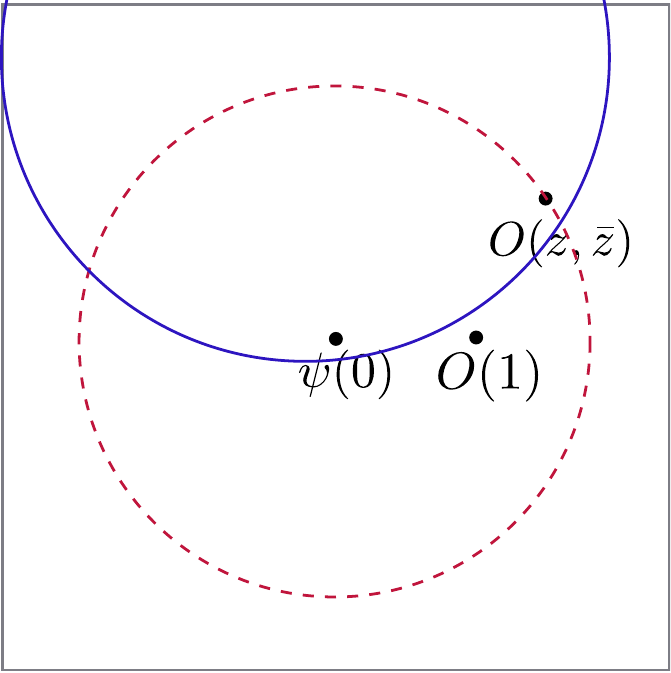}
\caption{\small The usual $\psi(0)O(z,\bz)$ OPE does not converge, since the dashed red circle contains another operator.  But if we expand around the origin of the solid blue circle, it converges. This is implemented by the $\rho$ variable. \label{fig:newcircle}}
\end{figure}

The $z$ expansion in the $s$ channel diverges for $|z|>1$. However we can still use the $\psi(0)O(z,\bz)$ OPE in this case by first mapping to a different configuration with the same value of the cross ratios.  It is clear that this should be possible for generic insertions, because if we picked a different origin of the $z$-plane instead of $z=0$, we could always find a circle that encloses $\psi(0)$ and $O(z,\bz)$ without hitting any other operators, see figure \ref{fig:newcircle}. Choosing the middle of this circle as the origin for radial quantization will give a convergent expansion.

 To achieve this explicitly, following \cite{Pappadopulo:2012jk,Hogervorst:2013sma}, insert the operators as
\be\label{defh}
H(\rho, \brho) \equiv \langle \psi(-\rho) O(\rho) O(1)\psi(-1) \rangle , 
\ee
where $\rho$ is a complex number with $|\rho| <1$.  Here we are labeling points in $R^d$ by the complex coordinate $y + i \tau$, with other directions $x^i=0$. The cross ratio for this configuration, and its inverse, are
\be\label{rhodef}
z(\rho) = \frac{4\rho}{(\rho+1)^2} , \qquad \rho(z) = \frac{z}{(1 + \sqrt{1-z})^2} \ ,
\ee
and similarly for $\bz, \brho$. The corresponding coordinate change is
\be
w'(w) = \frac{2(\rho+w)}{(\rho+1)(w+1)}\ ,
\ee
mapping operators inserted at $w=(-\rho,\rho,1,-1)$  to $w' = (0,z,1,\infty)$. A circle around the $w$ origin maps for example to the shifted circle in the $w'$ plane shown in figure \ref{fig:newcircle}. Using the $\rho$ variable is a way of automatically choosing the origin for radial quantization in a way that avoids other operator insertions.

\begin{figure}
\centering
\includegraphics{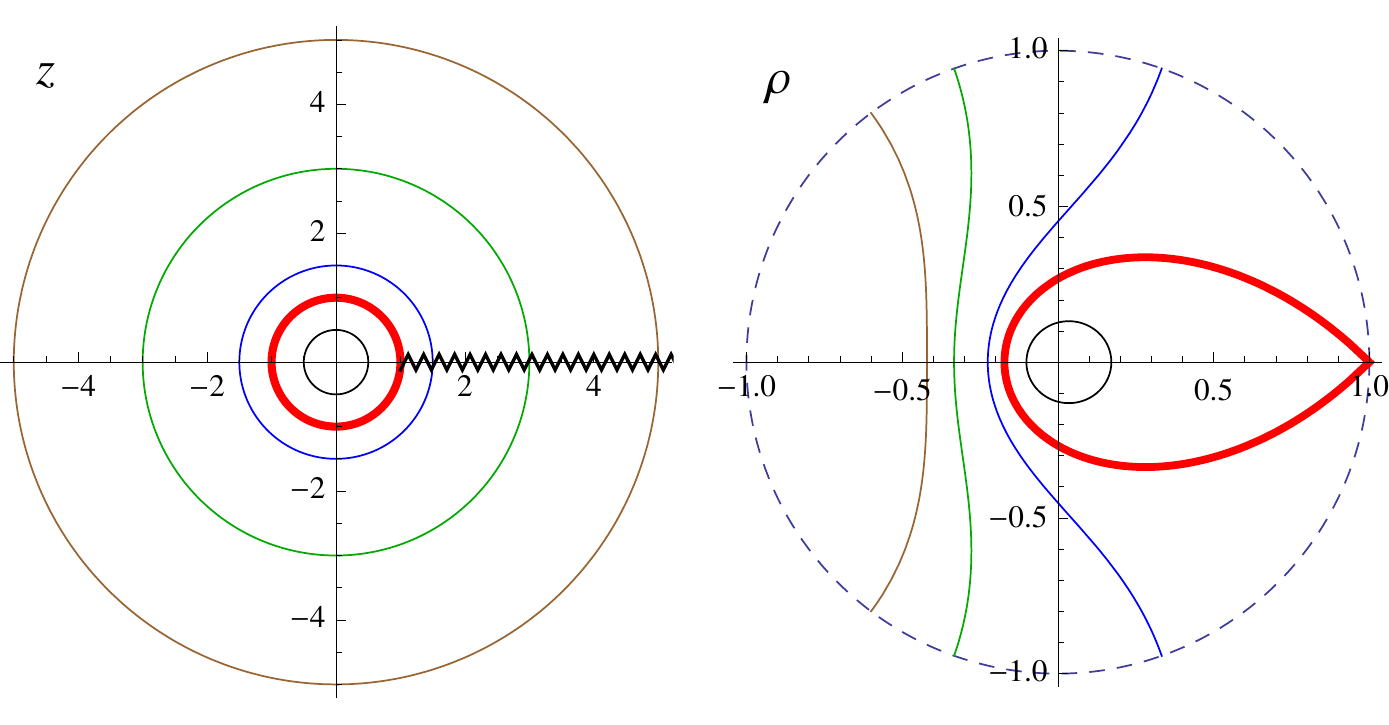}
\caption{\small \textit{Left:} Circles on the $z$ plane. \textit{Right:} Corresponding paths on the $\rho$ plane.  The thick red path, in both plots, is $|z|=1$. The branch cut along $[1,\infty]$ in the $z$ plane maps to $|\rho| = 1$.\label{fig:zrho}}
\end{figure}

Using \eqref{genblocks} to expand $H(\rho, \brho)$ in the $s$ channel implies
\be\label{srho}
G(z,\bz) = \left( z \bz \over 16 \rho \brho\right)^{-\half(\Delta_O+\Delta_\psi)} \left[ (1+\rho)(1+\brho)\over (1-\rho)(1-\brho) \right]^{-\Delta_{\psi O}} H(\rho, \brho) \ ,
\ee
with $\rho = \rho(z), \brho = \rho(\bz)$. The  $\psi(-\rho) O(\rho)$ OPE converges inside the 4-point function for any $|\rho|<1$, which maps to the full $z$ plane, minus the line $[1,\infty]$. This is shown in figure \ref{fig:zrho}, with the curve $|z|=1$ in red. Expanding the the rhs of \eqref{srho} as a power series in $\rho,\brho$ gives an expansion of $G(z,\bz)$ that converges on this larger domain.

 Each channel has a corresponding $\rho$-expansion: $\rho(z)$, $\rho(1-z)$, and $\rho(1/z)$.  Therefore, by mapping to the $\rho$ variable, we can always expand any Euclidean correlator in all three channels, $s$, $t$, and $u$ (unless all four points are colinear).   As power series expansions in $\rho$, all three channels converge for Euclidean $z \in {\mathbf C} \backslash [1,\infty]$.

\subsection{Positive coefficients}\label{ss:positivity}

\subsubsection*{Positivity in the $z$ expansion}
The $s$ channel is a sum of non-negative powers of $z$ and $\bz$, times a prefactor:
\be\label{defahh}
G(z,\bz) = (z\bz)^{-\half(\Delta_O + \Delta_\psi)} \sum_{h, \bh \geq 0} a_{h, \bh} z^h \bz^{\bh} \ .
\ee
The exponents are $h = \half(\Delta \pm \ell)$, $\bh = \half(\Delta \mp \ell)$, and the sum is over all $\Delta,\ell$ in the $O \psi$ OPE (not just primaries).
We will now show, using reflection positivity, that this expansion has positive coefficients,
\be\label{apos}
a_{h,\bh} \geq 0 \ .
\ee
The coefficients in the $u$ channel are identical, so they are also positive. These facts will be useful to bound the magnitude of the correlator in various regimes, and to understand when the expansion converges in Lorentzian signature, where $z$ and $\bz$ are independent. The derivation is based on a similar result for individual conformal blocks in \cite{Fitzpatrick:2012yx}.

Define a state (in radial quantization) by smearing the $O$ insertion over the Euclidean unit disk:
\be\label{jmstate}
|f\rangle \equiv \int_0^1 dr_1 \int_0^{2\pi} d\theta_1 \, f(r_1,\theta_1) O(r_1e^{i\theta_1}, r_1 e^{-i \theta_1})\psi(0)|0\rangle \ ,
\ee
where $f(r,\theta)$ is some smooth function.  
Reflection positivity is the condition $\langle f | f \rangle > 0$. In radial quantization,  conjugation acts on operators by inversion across the unit sphere,  
\be
[O(z,\bz)]^\dagger = (z\bz)^{-\Delta_O}O^\dagger(\frac{1}{\bz}, \frac{1}{z}) \ ,
\ee 
so reflection positivity imposes a condition on the doubly-integrated 4-point function.  (This notation is usually used in 2d CFT, but since all our operators are restricted to a plane, it is the same here.) In appendix \ref{app:positive} we show that applying this for all $f$ implies \eqref{apos}.

In $d=4$, this can also be checked order by order using the explicit conformal block derived by Dolan and Osborn \cite{Dolan:2000ut} and reproduced in appendix \ref{app:doblocks}, though it requires some subtle matching of conventions between blocks and OPE coefficients. Up to an overall normalization (that differs among references), the conformal blocks themselves have an expansion with positive coefficients when $\Delta_{12} = -\Delta_{34}$:
\be
g_{\Delta_p, \ell_p}^{\Delta_{12}, -\Delta_{12}}(z,\bz) = (-1)^{\ell_p} z^{-a}\bz^{-b} \sum_{p,q \in Z_+} (positive) \times z^p \bz^q \ .
\ee
This can be checked to any desired order by expanding the hypergeometric functions (and  is proved in appendix A.2 of \cite{Fitzpatrick:2012yx} for $\Delta_{12}=0$). We have picked the normalization in which OPE coefficients for scalar operators are real, as in \cite{Kos:2014bka}. The 3-point coefficients obey $c_{O_1 O_2 p}  = c_{O_2 O_1 p}(-1)^{\ell_p}$, so the $(-1)^{\ell_p}$ factors cancel in \eqref{sblocks}, confirming that the full correlator has an expansion with positive coefficients. This is also checked for two decoupled scalars in appendix \ref{app:positive}.

\subsubsection*{In the $\rho$ expansion}
A similar argument shows that $H(\rho, \brho)$ has an expansion in $\rho$ with positive coefficients:
\be
H(\rho, \brho) = (16 \rho \brho)^{-\half(\Delta_O + \Delta_\psi)} \sum_{h, \bh\geq 0}b_{h, \bh}\rho^h \brho^{\bh} , \qquad b_{h,\bh}>0 \ .
\ee
We omit the details. Using \eqref{srho}, it follows that $G(z,\bz)$ can be expanded $\rho$ with positive coefficients, after stripping off the correct prefactor:
\be\label{zrhoe}
G(z,\bz) = (z\bz)^{-\half(\Delta_O + \Delta_\psi)} \left[ (1+\rho(z))(1+\brho(z))\over (1-\rho(z))(1-\brho(\bz))\right]^{-\Delta_{\psi O}} \sum_{h, \bh\geq 0}b_{h, \bh}\, \rho(z)^h \brho(\bz)^{\bh} \ .
\ee
There is no obvious connection between the conditions $a_{h,\bh} > 0$ and $b_{h, \bh} > 0$.

As a check, we can also confirm this in $d=4$ using the Dolan-Osborn blocks. By explicit computation to high order, one indeed finds that the combination
\be
(-1)^{\ell_p} \left[ (1+\rho)(1+\brho)\over (1-\rho)(1-\brho)\right]^{\delta} g_{\Delta_p, \ell_p}^{\delta, -\delta}(z(\rho), \bz(\brho)) \ ,
\ee
where $g$ is the Dolan-Osborn block in $d=4$, has an expansion in $\rho,\brho$ with positive coefficients. (This is not true without the prefactor.)

\subsection{Lightcones in the $t$ channel $O \to O$}\label{ss:t}

\subsubsection*{Basics}
So far, we have mostly discussed the $s$ and $u$ channels, in which $O \to \psi$.  These are the channels where the expansion has positive coefficients. In the $t$ channel \eqref{tblocks}, $O \to O$, the coefficients can have either sign. Permutation symmetry $c_{12p} = c_{21p}(-1)^{\ell_p}$ implies that only even spins appear in this channel for real external fields. 

The leading term as $z,\bz\to1$ is the identity contribution, $[(1-z)(1-\bz)]^{-\Delta_O}$.  How the corrections are organized depends on how we take the limit.  If we take the limit $z\to 1, \bz \to 1$ at the same rate, so $(1-z)/(1-\bz)$ is held fixed, then \eqref{tblocks} is an expansion in conformal dimension, with the leading corrections coming from the operator of lowest $\Delta$.%

If we instead take the lightcone limit $\bz \to 1$ with $z$ held fixed, then \eqref{tblocks} is an expansion in twist $\Delta - \ell$,
\be\label{twistex}
G(z, \bz) = [(1-z)(1-\bz)]^{-\Delta_O}(1 + \lambda_m(1-\bz)^{\half (\Delta_m - \ell_m)} \tilde{g}_{\Delta_m, \ell_m}(1-z) + \cdots ) \ ,
\ee
where $\Delta_m$ is the dimension of the operator $O_m$ with minimal twist in this channel, $\ell_m$ is its spin, and 
\be
\lambda_m = c_{OO O_m}c_{\psi\psi O_m} \ .
\ee
$\tilde{g}$ in \eqref{twistex} is a known function called the lightcone (or colinear) conformal block.  Its explicit form is discussed below.  Corrections to \eqref{twistex} are suppressed by positive powers of $(1-\bz)$. We have assumed a single operator with minimal twist, and $\Delta_m - \ell_m > 0$, so this excludes 2d CFT where all Virasoro descendants of the vacuum have twist zero. We will assume throughout the paper that $d>2$ and that there is a single operator of minimal twist, unless stated otherwise.

\subsubsection*{Lightcone blocks}

The function $\tilde{g}$ is the lightcone conformal block,
\be\label{lcblockf}
\tilde{g}_{\Delta, \ell}(w) = (-2)^{-\ell} w^{\half(\Delta+\ell)}\, _2F_1(\half(\Delta+\ell), \half(\Delta+\ell), \Delta+\ell, w) \ .
\ee
It is related to the ordinary conformal block by
\be
g_{\Delta, \ell}^{0,0}(1-z,1-\bz) = (1-\bz)^{\half(\Delta - \ell)} \tilde{g}_{\Delta,\ell}(1-z)+...
\ee
as $\bz \to 1$ with $z$ held fixed. \eqref{lcblockf} holds in any number of dimensions, whereas the full block $g$ is known in closed form only in even dimensions. 

The lightcone block \eqref{lcblockf} has a branch cut along $w \in (1,\infty)$. For example, plugging in the dimension and spin corresponding to a 4d stress tensor,
\be\label{tlblock}
\tilde{g}_{4,2}(1-z) = -\frac{15(3-3z^2 + (1+4z+z^2)\log z)}{2(1-z)^2} \ .
\ee
Going around the cut sends $\log z \to \log z \pm 2 \pi i$ in this expression.  In general, the behavior around the cut can be obtained from
\be
_2F_1(h,h,2h,1-ze^{-2\pi i}) = \,_2F_1(h,h,2h,1-z) + 2 \pi i \frac{\Gamma(2h)}{\Gamma(h)^2} \,_2F_1(h,h,1,z) \ .
\ee
For future reference, the leading term if we go around the cut and then approach $z\sim 1$ is:
\be
\tilde{g}_{\Delta,\ell}(1-ze^{-2\pi i}) \approx 2\pi i (-2)^{-\ell}(1-z)^{1 - \half(\Delta+\ell)}\frac{\Gamma(\Delta+\ell)^2}{(\Delta+\ell-1)\Gamma(\half(\Delta+\ell))^4} \ .
\ee

\subsubsection*{When is the $t$ channel reliable?}
The $t$-channel converges absolutely for independent $z$ and $\bz$ with $|1-z| < 1$, $|1-\bz| < 1$. This does not follow immediately from the Euclidean expansion, as it did in the $s$ channel, since the coefficients are not positive.  Instead we note $|c_{OOp}c_{\psi\psi p}| \leq \half (|c_{OOp}|^2 + |c_{\psi\psi p}|^2)$, so absolute convergence of $\langle OOOO\rangle$ and $\langle \psi\psi\psi\psi\rangle$ in the $t$ channel imply the same for $\langle \psi OO \psi\rangle$.

We will use the $t$ channel in two limits: $|z| \sim |\bz| \sim 1$, and the lightcone limit $|\bz|\to 1$ with fixed $z$. The $t$ channel is reliable in both of these limits, and can be used to state some bounds that will be useful later.

If both $|z| \to 1$ and $|\bz| \to 1$, the corrections in the $t$ channel are dominated by a single term $\sim (1-z)^{h_*}(1-\bz)^{\bh_*}$, with $h^*, \bh^* \geq \half(\Delta_m - \ell_m)$.  Therefore in some neighborhood of $|z|=|\bz| = 1$, we can bound
\be\label{lb}
|G(z,\bz) | \leq |(1-z)(1-\bz)|^{-\Delta_O}(1 + \mbox{const.} \times |1-z|^{\half(\Delta_m-\ell_m)}  |1-\bz|^{\half(\Delta_m-\ell_m)} ) \ ,
\ee
for some real constant. \eqref{lb} holds for any way of taking the limit; we have not assumed anything about the ratio $(1-z)/(1-\bz)$.

Now consider the lightcone limit $\bz \to 1$ with $z$ held fixed.  Denote the arbitrary fixed value of $z$ by $z=z_0$.  In this limit, we expect that \eqref{twistex} is reliable unless $z_0$ is a singularity, \ie as long as $[(1-z)(1-\bz)]^{\Delta_O}G(z,\bz)$ is regular as $\bz \to 1, z \to z_0$.  This is an important caveat. It means that we cannot use the lightcone expansion in the $t$ channel to show that the correlator is regular as a function of $z$, for example to test for causality.  Using this expansion already assumes regularity.\footnote{For example, consider the function $f(z,\bz) = 1 - \bz + \frac{(1-\bz)^2}{z-a}$. If we write $f(z,\bz) = 1 -\bz + O((1-\bz)^2)$ we might falsely conclude the function is regular at $z=a$.}

With this caveat, the lightcone expansion \eqref{twistex} is also reliable after analytically continuing $z \to e^{\pm 2\pi i }z$. This seems clear from the form of the $t$ channel sum, but we do not know of a direct proof,
so instead give an indirect argument using crossing symmetry and the results of \cite{Komargodski:2012ek,Fitzpatrick:2012yx}. These papers showed that $s=t$ crossing symmetry, in the lightcone limit, fixes the dimensions of certain high-spin operators in the $s$ channel.  Roughly speaking they solved the equation
\be\label{dimcr}
\sum_{h,\bh} a_{h,\bh} z^h \bz^h \approx \frac{[(1-z)(1-\bz)]^{-\Delta_O} }{(z\bz)^{-\half(\Delta_O+\Delta_\psi)}}\left(1 + \lambda_m(1-\bz)^{\half (\Delta_m - \ell_m)} \tilde{g}_{\Delta_m, \ell_m}(1-z) \right)
\ee
viewed as an equation for the spectrum $h,\bh$ and the OPE coefficients $a_{h,\bh}$. Given this solution, we can continue to the second sheet by evaluating  $\sum_{h,\bh} a_{h,\bh}z^h \bz^{\bh} e^{-2\pi i h}$.  But this will clearly agree with simply taking $z \to e^{-2\pi i}z$ on the rhs of \eqref{dimcr}. 
Therefore, there is a subsector of operators whose contribution in the $s$ channel produces the analytically continued $t$ channel answer on the second sheet. It remains to show that this subsector dominates the correlator; in the regime of interest we will do this in section \ref{ss:logst},  confirming that we can trust the lightcone limit of the $t$ channel on the second sheet.

\subsection{Causality in a Simple Case}\label{ss:simplecausality}
We now specialize to real $z,\bz$ and given an example where we can directly relate reflection positivity to causality.  We will use only the $s$ channel in this example.

\begin{figure}
\centering
\includegraphics{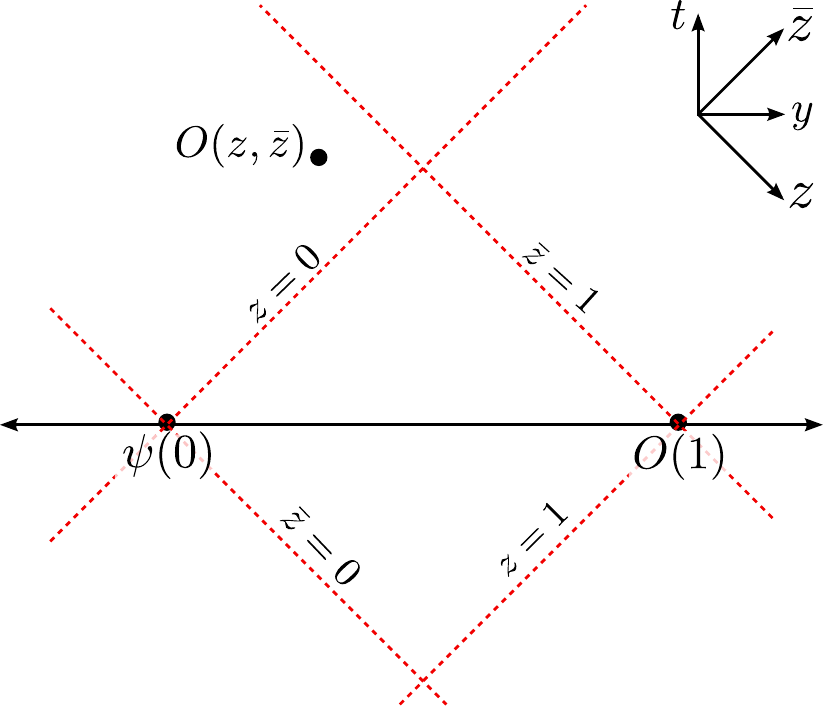}
\caption{ \small Insertion points in the Lorentzian 4-point function. The 4th insertion $\psi(t=0, y=\infty)$ is not shown. \textit{Caveat:} This picture does not apply to the shockwave kinematics discussed later in the paper.\label{fig:zbz}}
\end{figure}

In Euclidean signature, $\bz = z^*$, but in Lorentzian signature $z$ and $\bz$ are independent. The simplest case is the Lorentzian correlator $\langle \psi(0) O(z,\bz) O(1) \psi(\infty)\rangle$ where now $z$ and $\bz$ are independent real numbers,
\be
z = y_2 - t_2 , \qquad \bz = y_2 + t_2 \ .
\ee  
This configuration is shown in Lorentzian signature in figure \ref{fig:zbz}.  Three of the insertions are spacelike separated, so we only need to worry about how $O(z,\bz)$ is ordered with respect to the other operators.

The first question is when the OPEs converge inside this Lorentzian correlator. The $s$ channel converges for Euclidean $|z| < 1$.  Since the coefficients of this expansion are all positive, the $s$ channel also converges for independent complex $z$ and $\bz$ satisfying
\be\label{radcon}
|z| < 1 , \quad |\bz| < 1 \ .
\ee
This follows from writing the $s$ channel sum as $\sum_{h,\bh} a_{h,\bh} |z|^h |\bz|^{\bh} e^{i h \phi - i \bh \bphi}$ where $\phi, \bphi$ are real phases. Since $a_{h,\bh} > 0$ these phases can only decrease the magnitude of the sum, and it must converge.

In particular, the $s$ channel converges in the timelike-separated configuration of fig.~\ref{fig:zbz}, with $-1<z<0$ and $0<\bz<1$. We will now use this expansion to prove that this correlator is causal in the sense
\be\label{simplec}
\langle \psi(0)\, [\, O(z,\bz), O(1)\, ]\,  \psi(\infty)\rangle = 0 \qquad (0<\bz<1, \quad z<1) \ ,
\ee
including negative $z$ where $\psi$ and $O$ are timelike separated. 
First, note that this is obvious for $z>0$, \ie if $\psi$ is spacelike separated from the $O$'s.  In this case, we can expand in the convergent $t$ channel, and the leading behavior $[(1-z)(1-\bz)]^{-\Delta_O}(1+\cdots)$ shows that the singularity is precisely on the lightcone.  

It is nontrivial for $z<0$.  Fix $0<\bz <1$, and define $x = |z| \in (0,1)$. Consider the change in the correlator as we reflect across the $\psi$ lightcone by sending $z=x$ to $z = -x$, as shown here:
\be
\begin{gathered}\includegraphics{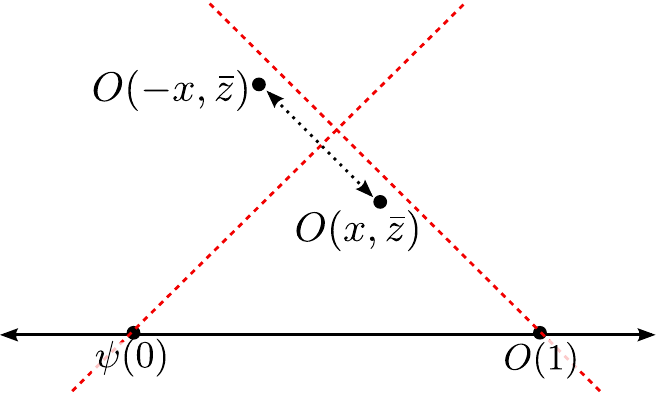}\end{gathered} \ .
\ee
The positive expansion in the convergent $O(z,\bz) \to \psi(0)$ channel implies
\be\label{cinq}
\big|\langle \psi(0) O(-x,\bz)O(1)\psi(\infty) \rangle \big| = |G(xe^{-i\pi},\bz)| \leq |G(x, \bz)| 
\ee
and
\be\label{cinq2}
\big|\langle O(-x,\bz)\psi(0) O(1)\psi(\infty) \rangle \big|= |G(xe^{i\pi},\bz)| \leq |G(x, \bz)|  \ .
\ee
That is, the correlator can only \textit{decrease} in magnitude as we reflect $O(x,\bz)$ across the $\psi(0)$ lightcone.   

To diagnose causality, as described in section \ref{ss:cftcausality}, we send $\bz \to 1$ with fixed $z<0$, and look for the singularity.  The inequality \eqref{cinq} means that the singularity as a function of $\bz$ (called $\bz_0$ in section \ref{ss:cftcausality}) cannot shift to earlier times on the second sheet of the analytically continued correlation function.  This implies \eqref{simplec}, in the range $-1<z<1$. To recap: This commutator becomes non-zero when $G(z, \bz)$ hits a singularity as a function of $\bz$. This does not happen for $0<z<1$ (since the correlator is obviously causal on the first sheet), so according to \eqref{cinq} it cannot happen for $-1<z<0$, either. 

The argument fails for $z<-1$, since the $z$ expansion may diverge. But in this case we can turn to the $\rho$ expansion: $\sum b_{h,\bh}\rho(z)^h \brho(\bz)^{\bh}$
converges for $|\rho| < 1$, $|\brho|<1$, and this range includes all real $z<1$.

Therefore we have shown, in this simple case of a single timelike separation, how reflection positivity is linked to causality.  This simple argument did not use crossing symmetry beyond the initial choice of the $s$ channel, and we do not expect it to lead to any useful new bounds on CFTs beyond the obvious constraints on three-point functions required by reflection positivity.

\section{Causality of Shockwaves}\label{s:cshock}

\begin{figure}
\centering
\includegraphics{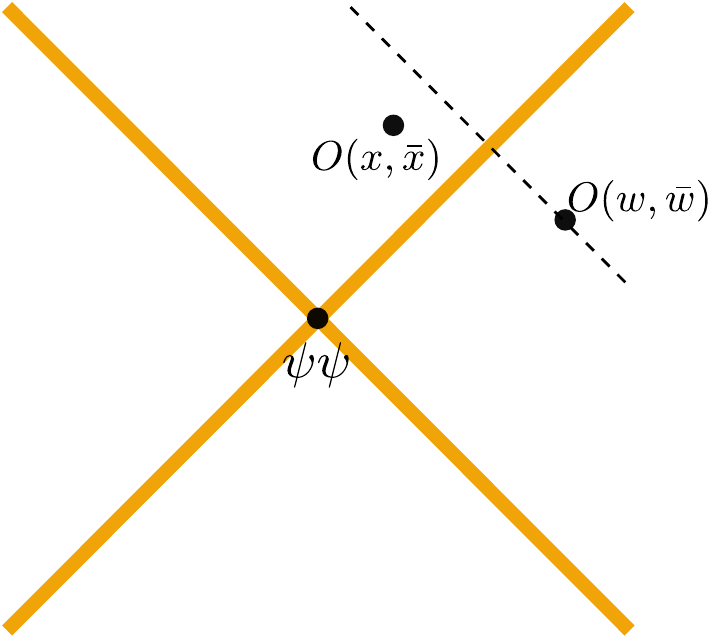}
\caption{\small Kinematics of the shockwave 4-point function.  The two $\psi$'s are inserted at $t = \pm i \delta$, $\vec{x} = 0$, so the shockwave is spread over a width $\sim 2\delta$. \label{fig:shocksetup}}
\end{figure}

We now turn to causality bounds for more complicated Lorentzian configurations, where two operators are timelike separated from the others.  These are closely related to expectation values,
\be\label{oocom}
\langle \Psi | [O, O] |\Psi \rangle \ .
\ee
This must be causal in any state $|\Psi\rangle$.  Let us choose the state
\be
|\Psi \rangle \equiv (2\delta)^{\Delta_\psi} \psi (-i\delta) |0\rangle \ ,
\ee
where $\delta > 0$. As above, points in Euclidean $R^d = (\tau, y, x^2, \dots x^{d-1})$ are labeled by the complex coordinate $y + i \tau$, with $x^{2,\dots,d-1} = 0$. The operator insertion is offset in Euclidean time in order to define a state in the Lorentzian theory with a finite norm, and the prefactor sets $\langle \Psi | \Psi \rangle = 1$.  We can roughly think of this state at $t=0$ as a lump of $\psi$ concentrated near the origin, with width $2\delta$.  Viewed on length scales much greater than $\delta$, the $\psi$ insertion creates a shockwave at the origin that propagates outward at the speed of light, as depicted in figure \ref{fig:shocksetup}. This means that we can think of the commutator \eqref{oocom} as a measure of whether the operator $O$ can be used to send a signal in the background of the $\psi$ shockwave. 

The comparison to gravitational shockwaves in AdS/CFT will be discussed in a future paper.  Our setup differs somewhat from the work of \cite{Cornalba:2006xk,Cornalba:2006xm} on shockwaves and the eikonal limit in AdS/CFT, in particular we do not assume large $N$, but ultimately we study a similar limit of the correlator.

\subsection{Regime of the 4-point function}

Causality of the $[O,O]$ commutator in this state is related to the analytic structure of the 4-point function
\bea
I(x,\bx, w, \bw) &\equiv& \langle \Psi |O(w, \bw)  O(x,\bx)| \Psi \rangle \notag\\
&=&(2\delta)^{2\Delta_\psi} \langle \psi(i\delta)O(w, \bw) O(x,\bx)  \psi(-i\delta)\rangle \ .\label{ixdd}
\eea
Although this 4-point function has both Euclidean and Lorentzian insertions, we emphasize that it computes a physical, real-time expectation value, that could be directly measured in an experiment. The insertion of $\psi$ at complex time simply prepares the initial state for this Lorentzian experiment.

We assume
\be\label{xwr}
x,\bx,w,\bw \in \mathbf{R} \ , \quad w>0, \qquad \bw > \bx \gg \delta > 0 
\ee
as in figure \ref{fig:shocksetup}.
Applying \eqref{genblocks} relates this expectation value to the canonical 4-point function,
\be
I = (4\delta^2)^{\Delta_O} [(w\bw+\delta^2)(x\bx + \delta^2)]^{-\Delta_O} (z\bz)^{\half(\Delta_\psi + \Delta_O)}G(z,\bz)
\ee
with the cross ratios
\be
z = \frac{(x-i\delta)(w+i\delta)}{(x+i\delta)(w-i\delta)} , \quad \bz = \frac{(\bx+i\delta)(\bw-i\delta)}{(\bx-i\delta)(\bw+i\delta)} \ .
\ee
Note that $|z| = |\bz| = 1$. Under the assumptions \eqref{xwr},
\be
\bz \approx 1 + i \bvep , \qquad \bvep \equiv 2\delta \left( \frac{1}{\bx} - \frac{1}{\bw} \right) \ .
\ee
The other cross ratio is also near 1 in the limit $\delta \to 0$,
\be
z = 1 - 2 i \delta \left( \frac{1}{x} - \frac{1}{w}\right) +O(\delta^2) \ ,
\ee
but this expression breaks down for small enough $x$ or $w$ so we need to be careful as we cross the shockwave.

Following the logic of section \ref{s:causality}, to find the correlator $I(x,\bx,w,\bw)$ past the shockwave, we need to specify a path in the complex-$\tau_2$ plane that selects the correct operator ordering.  How to do this was explained briefly in \cite{Asplund:2014coa} and will now be repeated in the present notation. Fix $w,\bw$ and $y_2 = \frac{1}{2}(x +\bx)$.  The analytic structure as a function of complex $\tau_2 = \frac{1}{2i}(x - \bx)$ is:
\be\label{f:pathbetween}
\includegraphics{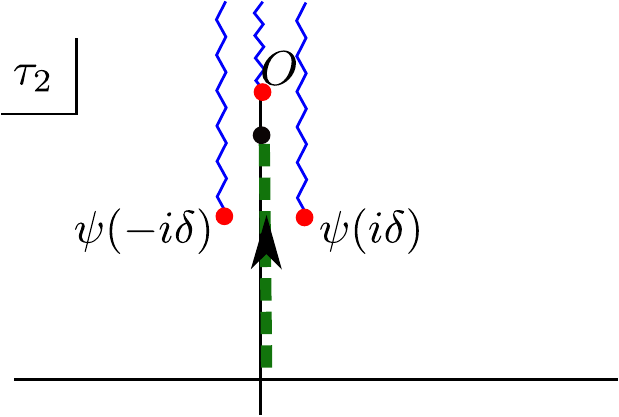}
\ee
To compute the correlator ordered as in \eqref{ixdd}, we want to time-order $O(x,\bx)$ with respect to $\psi(-i\delta)$, and anti-time order it with respect to $\psi(i\delta)$. Therefore we analytically continue along a path that goes to the left of $\psi(i\delta)$ and to the right of $\psi(-i\delta)$, as drawn (green dashed line).

Causality requires that for fixed $x<w$, the singularity in $I$ as a function of $\bx$ does not appear before $\bx = \bw$.  That is, the singularity as $O(x,\bx)$ approaches the $O(w,\bw)$ lightcone cannot appear too soon. This holds trivially for $0<x<w$ because this is the first sheet of the analytically continued Euclidean correlator, but is nontrivial for $x<0$, \ie after $O(x,\bx)$ has passed through the shockwave.

\begin{figure}
\centering
\includegraphics{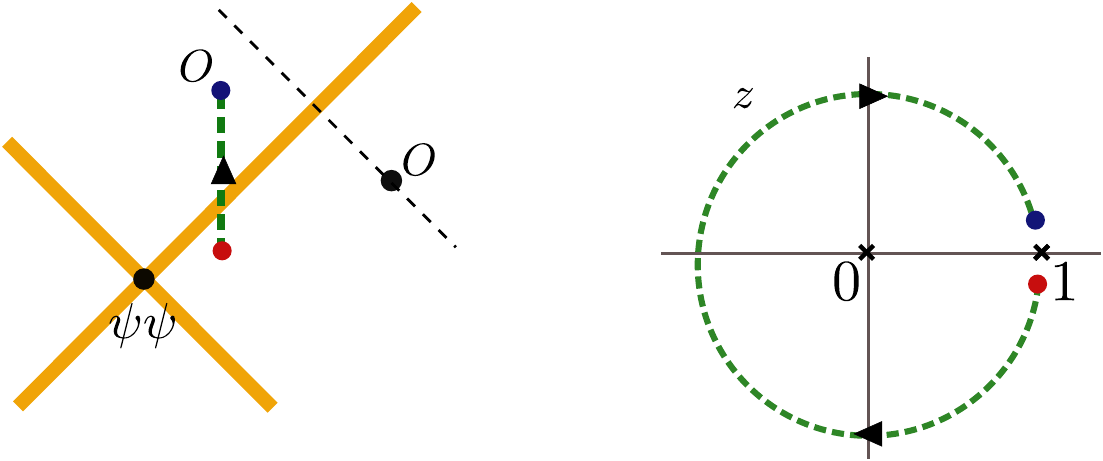}
\caption{The analytic continuation appropriate to the ordering $\psi OO \psi$ is the dashed green line in \eqref{f:pathbetween}.  This same path is shown here in real spacetime (\textit{left}) and on the $z$ plane (\textit{right})..\label{fig:zcontour}}
\end{figure}

What does causality mean for $G(z,\bz)$? To rephrase this in the $z$ variable, we follow $z,\bz$ as $O(x,\bx)$ goes through the shockwave, along the path in \eqref{f:pathbetween}.
Away from the shock, for $|x|, w \gg \delta$, we have $z \approx \bz \approx 1$.  But as we cross the shock at $x=0$, heading in the positive-$t$ (\ie negative-$x$) direction, the cross-ratio $z$ circles clockwise around the unit circle. This is shown in figure \ref{fig:zcontour}. $\bz$ does nothing interesting; it just stays near 1. So the conclusion (up to prefactors) is that
\be
G(ze^{-2\pi i}, \bz) \propto  \langle \psi(i\delta)O(w, \bw) O(x,\bx)  \psi(-i\delta)\rangle
\ee
after the shock, with operators ordered as written. We will refer to $G(ze^{-2\pi i}, \bz)$ as the correlator evaluated on the `second sheet.' For $|z|, |\bz| < 1$, it can be computed by the convergent $s$ channel expansion,
\be\label{gsumdef}
G(ze^{-2\pi i}, \bz) = (z\bz e^{-2\pi i})^{-\half(\Delta_O+\Delta_\psi)}\sum_{h,\bh\geq 0} a_{h,\bh} z^h \bz^{\bh} e^{-2\pi i h} \ ,
\ee
where the positive OPE coefficients $a_{h,\bh}$ were defined in \eqref{defahh}.
Outside this range, the notation $G(ze^{-2\pi i}, \bz)$ is shorthand for the function defined by analytic continuation of \eqref{gsumdef}. As long as we do not cross any singularities this is unambiguous.

To test for causality we send $O(x,\bx)$ near the $O(w,\bw)$ lightcone, $\bx \to \bw$. This is the limit $\bz \to 1$ from the positive-imaginary direction. Also assume $-x \gg \delta$, so $z$ is also near 1 in the upper half plane, but on the second sheet. Therefore, the commutator \eqref{oocom} is causal if and only if 
\be
 G(e^{-2\pi i}(1+i\ep), \, 1 + i \bvep)
\ee
is regular for 
\be\label{epdom}
1 \gg \ep, \bvep > 0 \ .
\ee
The simultaneous limits $\ep \to 0, \bvep \to 0$ can be taken in different ways, corresponding to different null limits of the correlator.  For example, if we hold fixed $\delta\ll 1$, set $w=\bw=1$, and send $O(x,\bx)$ to the light cone of $O(w,\bw)$, this is the limit $\bvep \to 0$ with $\ep$ small and fixed. On the other hand if take $\delta \to 0 $ first, then this is the limit $\ep \sim \bvep \to 0$. In both cases, causality requires the function is regular for all $\ep,\bvep > 0$.

In the rest of this section, we will show that this is the case using the OPE methods of section \ref{s:lope}.

\subsection{Bound from the $s$ channel}\label{ss:rhobound}
We will bound the magnitude of the correlator using the fact that the $\rho$ expansion has positive coefficients, derived in section \ref{ss:positivity}.  In terms of the correlator \eqref{defh}, positive coefficients imply
\be\label{rhoinh}
|H(\rho, \brho)| \leq H(|\rho|, |\brho|) \qquad \mbox{for} \quad |\rho|<1 , \quad |\brho| < 1 \ .
\ee
This bound, rewritten in the $z$ variable, implies
\be
|G(z,\bz)| \leq W(z,\bz)G(r(z), r(\bz))
\ee
where
\be
W(z,\bz) \equiv \left| \frac{r(z)r(\bz)}{z\bz}\right|^{\Delta_O} \left| \frac{(1-\rho(z))(1 - \rho(\bz))}{(1 - |\rho(z)|)(1-|\rho(\bz)|)}\right|^{\Delta_\psi - \Delta_O}
\ee
and, with the definitions in \eqref{rhodef},
\bea\label{rhoinz}
r(z) &\equiv& z( | \rho(z) | )\notag\\
&=& \frac{4|z||1 + \sqrt{1-z}|^2}{(|z| + |1 + \sqrt{1-z}|^2)^2} \ .
\eea
(Note  $|z| \equiv \sqrt{z z^*} \neq \sqrt{z \bz}$). This bound holds for any complex $z$, as long as we do not cross through $|\rho|=1$, so in particular it applies on the second sheet,
\be\label{rhoin}
|G(ze^{-2\pi i}, \bz)| \leq W(z,\bz)G(r(z), r(\bz)) \ .
\ee
This bounds the magnitude of $G(z,\bz)$ with complex arguments on the second sheet, in terms of $G(z,\bz)$ with real arguments on the first sheet.    Specializing to real $z,\bz$, it implies
\be\label{realb}
|G(ze^{-2\pi i}, \bz)| \leq G(z,\bz)  \quad \mbox{for} \quad z, \bz \in (0,1) \ .
\ee
\eqref{realb} is also an obvious consequence of positive coefficients in the $z$ expansion, but off the real line, the bound from positive $\rho$ coefficients is stronger.

More generally, we will apply \eqref{rhoin} in the domain
\be\label{lhpdomain}
\mbox{Im\ } z \geq 0 , \quad \mbox{Im\ } \bz \geq 0 , \quad 0 < \mbox{Re\ } z , \mbox{Re\ } \bz \leq 1 \ .
\ee
The actual range where the inequality holds is larger but requires care with the definition of $\sqrt{1-z}$ and will not be needed. For $z,\bz \approx  1$ in the domain \eqref{lhpdomain}, 
\be
r(z) \approx 1 - (\mbox{Re}\sqrt{1-z})^2  \ ,
\ee
and $W(z,\bz) \approx 1$. 
Therefore, invoking the $t$ channel expansion \eqref{tblocks},
\be\label{rbf}
|G(ze^{-2\pi i}, \bz)| \leq  \left[ (\mbox{Re}\sqrt{1-z}) (\mbox{Re}\sqrt{1-\bz}) \right]^{-2\Delta_O} (1 + \cdots  )\ ,
\ee
for $z \approx 1$, $\bz \approx 1$. Corrections are suppressed by positive powers of $1-z$ and $1-\bz$, organized by dimension or by twist, depending on the relative limits.  Note that it is the $t$ channel expansion on the \textit{first} sheet that appears on the right-hand side of \eqref{rbf}. This ensures  the corrections have only positive powers.

In the domain \eqref{lhpdomain}, 
\be
\mbox{Re}\sqrt{1-z} = \cos(\half \arg (1-z))\sqrt{|1-z|} \leq \frac{1}{\sqrt{2}} \sqrt{|1-z|} \ ,
\ee
so for example on the plane Re $z = $ Re $\bz = 1$ we find
\be\label{imb}
|G(e^{-2\pi i }(1 + i \ep), 1 + i \bvep)| \leq G(1 - \half \ep, 1 - \half \bvep) \ , 
\ee
for $ 0 < \ep,\bvep \ll 1$. 

\subsection{Causality}
As explained above, the relevant limit for causality of the shockwave is $z = 1  + i \ep$, $\bz = 1 + i \bvep$ on the second sheet, with real $\ep, \bvep \ll 1$.  In this limit, the bound \eqref{imb} allows the correlator to grow no faster than the singularity on the first sheet, times a constant:
\be
|G(e^{-2\pi i}(1 + i \ep), 1 + i \bvep) | \leq 4^{\Delta_O} (\ep \bvep)^{-\Delta_O}(1+\cdots) \ ,
\ee
with corrections suppressed by positive powers of $\ep, \bvep$. This shows that there is no singularity for $\ep,\bvep >0$.  Therefore, the commutator $\langle \Psi | [O,O]|\Psi\rangle$ vanishes outside the lightcone.\footnote{A similar approach might be  useful to examine the unexpected singularities in CFT 4-point functions on a cylinder that appear in AdS/CFT as an artifact of the $1/N$ expansion, and are related to bulk scattering \cite{Heemskerk:2009pn}. See \cite{zhibtalk} for a discussion. }

\section{Constraints on Log Coefficients}\label{s:logs}

So far, we showed there was no acausal singularity on the shockwave background; now we will bound the finite corrections that appear just before the lightcone.
Let
\be
z = 1 + \sigma , \qquad \bz = 1 + \eta \sigma
\ee
with complex $\sigma$ and real $\eta$ satisfying
\be
 \quad \mbox{Im\ } \sigma \geq 0 \ , \quad |\sigma| \ll 1   \ , \quad 0<\eta \ll 1 \ .
\ee
The limit $\eta \to 0$ with fixed $\sigma$ is the lightcone limit, while $\sigma \to 0$ with fixed $\eta$ is the Regge limit.\footnote{A different relationship between the lightcone limit and Regge limit was explored perturbatively in \cite{Costa:2013zra}.} We will argue that in both cases (or any other limit with $\sigma, \eta \to 0$), the correlator on the second sheet is bounded by the identity in the $t$ channel $O \to O$ on the first sheet, up to small corrections:
\be\label{gzze}
|G(e^{-2\pi i }(1+\sigma), 1+\eta\sigma)| \leq |\eta\sigma^2|^{-\Delta_O}\left[1 + \mbox{const.}\times \eta^{\half(\Delta_m - \ell_m)}\right] \ .
\ee
Recall that $\Delta_m, \ell_m$ are the dimension and spin of the minimal-twist operator appearing in the $t$ channel. The important feature of this inequality is that the correction term has no negative powers of $\sigma$. This fixes the sign of the coefficients of certain log terms in the $t$ channel conformal block expansion.  

Put differently, we will show that if the log term appears with the wrong sign, then both causality and reflection positivity are violated.

We will first collect a few facts about the correlator on the second sheet, then put them together to compute the coefficient of the log in terms of manifestly positive OPE data.  The strategy to do this was sketched in section \ref{s:brief}.

\subsection{Analyticity}

\begin{figure}
\centering
\includegraphics{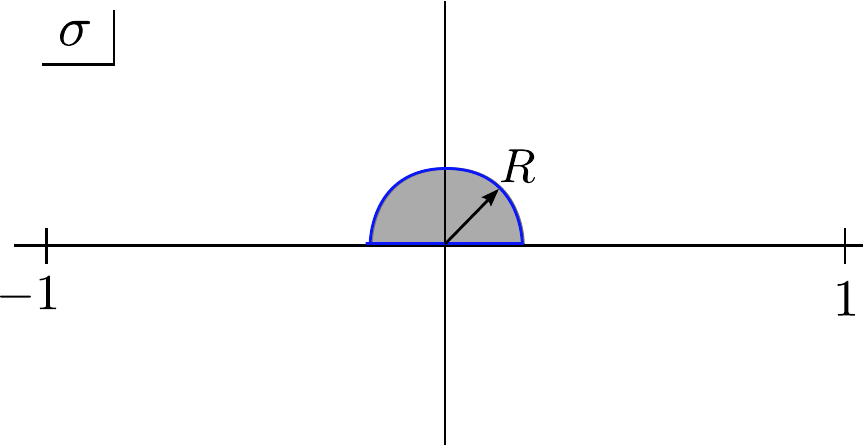}
\caption{\small Region $D$ in the complex $\sigma$ plane. The origin is excluded. \label{fig:d}}
\end{figure}

Define, on the first sheet,\footnote{\eqref{getafirst} has a branch cut at $\sigma=0$, so needs to be defined carefully. We are defining $G_\eta(\sigma)$ in such a way that it is equal to the Euclidean correlator for $\eta=1$, $\sigma \in R$.}
\be\label{getafirst}
G_\eta(\sigma) = (\eta\sigma^2)^{\Delta_O}G(1+\sigma, 1 + \eta \sigma)
\ee
and, on the second sheet,
\be
\widehat{G}_\eta(\sigma) = (\eta\sigma^2)^{\Delta_O}G(e^{-2\pi i}(1+\sigma), 1 + \eta \sigma) \ .
\ee
These are the normalized correlators with different orderings,
\bea
G_\eta(\sigma) &\sim& \frac{\langle O\psi O\psi\rangle}{\langle \psi \psi \rangle \langle OO \rangle } \ , \\
\widehat{G}_\eta(\sigma) &\sim&   \frac{\langle \psi O O\psi\rangle}{\langle \psi \psi \rangle \langle OO \rangle }  \ .
\eea
We are interested in the behavior of $\widehat{G}_\eta(\sigma)$ in the region $D$ shown in figure \ref{fig:d}, which is a small half-disk in the upper half $\sigma$ plane, with radius $R$ such that $\eta \ll R \ll 1$. The point $\sigma=0$ is excluded from $D$.

$\widehat{G}_\eta(\sigma)$ is analytic on region $D$ and finite at $\sigma=0$.  Analyticity follows from \eqref{rbf} and the similar bound from the $u$ channel: \eqref{rbf} implies that $|\widehat{G}_\eta(\sigma)| \lesssim 4^{\Delta_O}$ for $\mbox{Re}\, \sigma \leq 0$, and the positive $\rho$-expansion in the $u$ channel gives the same bound for Re $\sigma \geq 0$.\footnote{We are using the fact that the function obtained by sending $z \to e^{-2\pi i} z$ in the $s$-channel expansion is the same function, \ie on the same sheet of the multivalued correlator, as the function obtained by sending $\frac{1}{z} \to \frac{1}{z}e^{2\pi i}$ in the $u$ channel expansion.  This holds for Im $z,\bz \geq 0$ --- this can be checked by deforming the contours that define the analytic continuation --- but would not hold for Im $z,\bz <0$, where these two functions differ due to the choice of how to go around the singularity at 1. This is what restricts us to the upper-half $\sigma$ plane and ultimately what will be responsible for setting certain log coefficients to be positive rather than negative.
}

\subsection{Bound on the real line}\label{ss:rbb}
For real $\sigma \in [-R,0]$, the positive coefficients in the $s$-channel $z$ expansion imply that
\be\label{posrb}
|\widehat{G}_\eta(\sigma)| \leq G_\eta(\sigma)  \ , 
\ee
which combined with the $t$ channel result \eqref{lb} implies
\be
|\widehat{G}_\eta(\sigma)| \leq 1 + \mbox{const.}\times \eta^{\half(\Delta_m-\ell_m)} \ .
\ee
Since the coefficients in the $u$-channel $1/z$ expansion are also positive, these bounds also hold for $\sigma \in [0,R]$.

Also, the real part of the commutator $G_\eta(\sigma) - \widehat{G}_\eta(\sigma)$ is positive on the real $\sigma$ line.  For $\sigma \in [-R,0)$,
\begin{align}
\mbox{Re\ }&(G_\eta(\sigma) - \widehat{G}_\eta(\sigma)) \sim \mbox{Re\ } \frac{ \langle [O,\psi]O\psi\rangle}{\langle OO\rangle \langle \psi \psi \rangle}\\
&=\frac{ (\eta\sigma^2)^{\Delta_O}}{[(1+\sigma)(1+\eta\sigma)]^{\half(\Delta_O+\Delta_\psi)}} \sum_{\Delta, \ell} a_{\Delta, \ell}  r_{\Delta,\ell} (1+\sigma)^{\half(\Delta-\ell )} (1+\eta \sigma)^{\half(\Delta+\ell)}\notag\\
&\geq 0 \ ,\notag
\end{align}
where
$r_{\Delta,\ell} = 1 - \cos[\pi (\Delta-\ell - \Delta_O-\Delta_\psi)] $.
A similar expression for $\sigma \in (0,R]$ comes from the $u$-channel expansion. Up to corrections with only positive powers in $\eta,\sigma$, this can also be written
\be
\mbox{Re\ }\widehat{G}_\eta(\sigma) \lesssim 1 \ .
\ee

\subsection{Logs in $t$ channel}\label{ss:logst}
We have no expansion for $\widehat{G}_\eta(\sigma)$ around $\eta=\sigma=0$.  But in the lightcone limit $\eta \to 0$ with $\sigma\neq 0$ fixed, it was argued in section \ref{ss:t} that we can apply the $t$ channel expansion,\footnote{
\label{footnote:sa}
In section \ref{ss:t} we only showed that there is some subsector of operators producing \eqref{kvt}; now we also need to argue that this subsector dominates on the 2nd sheet in order to complete the argument that the $t$ channel is reliable in this regime. This follows from the bound in section \ref{ss:rhobound}: the contribution from other operators is enhanced by at most a multiplicative constant on the sheet, so the subsector of operators producing \eqref{kvt}, which dominate near the lightcone singularity on the first sheet, will also dominate on the 2nd sheet.}
\be\label{kvt}
 \widehat{G}_\eta(\sigma) = 1 + \lambda_m (-\eta\sigma)^{\half(\Delta_m -\ell_m)}\tilde{g}_{\Delta_m, \ell_m}(-\sigma) + \cdots \qquad (\eta \to 0) \ ,
 \ee
 where $m$ denotes the minimal-twist operator,
 \be
 \lambda_m = c_{OOm} c_{\psi\psi m} \ ,
 \ee
 and the lightcone block in \eqref{kvt} should be evaluated on the second sheet. 
We have assumed that there is a single operator of minimal twist, but this is readily generalized. Using the explicit expressions for the lightcone block $\tilde{g}$ in section \ref{ss:t}, the expansion for $\eta \ll \sigma \ll 1$ is
\be\label{logex}
\widehat{G}_\eta(\sigma) = 1 -  i \hat{\lambda}_m \frac{\eta^{\half(\Delta_m - \ell_m)}}{\sigma^{\ell_m-1}} + \cdots
\ee
where
\be\label{hatl}
\hat{\lambda}_m \equiv c_{OOm}c_{\psi\psi m}\times   \frac{2^{-\ell_m+1} \pi\Gamma(\Delta_m+\ell_m)^2}{(\Delta_m+\ell_m-1)\Gamma(\half(\Delta_m+\ell_m))^4} \ .
\ee
In \eqref{logex}, we are taking $\eta$ small enough so that the correction is small, thereby staying in the regime where the $t$ channel is reliable. Clearly this expression would be incorrect as $\sigma \to 0$ (at fixed $\eta$) since higher spin exchanges would dominate.

The key observation is that for $\ell_m>1$, the leading correction in $\widehat{G}_\eta(\sigma)$ has a negative power of $\sigma$.  There were no negative powers on the first sheet; they come only from the non-analyticity of the lightcone conformal block, after going around the branch cut, $\log z \to \log z - 2 \pi i$. The apparent `pole' $\sim \sigma ^{-\ell_m+1}$ is not actually singular, since $\widehat{G}_\eta(\sigma)$ is finite at $\sigma = 0$.  This is not a contradiction, since the expansion \eqref{logex} is for $\eta \ll |\sigma|$, and does not apply near the origin $|\sigma| \ll \eta$.

\subsection{Sum rule for log coefficient}
We will now put this all together to derive a formula for the log coefficient $\lambda_m = c_{OOm} c_{\psi\psi m}$ that is manifestly positive.  Analyticity implies
\be\label{contz}
 \oint_{\p D}d\sigma \, \sigma^k \widehat{G}_\eta(\sigma)  = 0\ ,
\ee
for any $k>-1$. The path $\p D$ consists of the real line segment $\sigma \in [-R, R]$ and the semicircle 
\be
S = \{ \sigma = R e^{i \phi}  \ , \phi \in [0, \pi] \} \ ,
\ee
oriented counterclockwise.
On the semicircle, the $t$ channel is reliable on both the first and second sheets, so we can do the integral explicitly be plugging in \eqref{logex}. Choosing $k = \ell_m - 2$
and using the identity (valid for integer $m$)
\be
\mbox{Re} \ i \int_S d\sigma \, \sigma^m = -\pi \delta_{m,-1} \ 
\ee
gives
\be\label{resec}
\mbox{Re}\ \int_S d\sigma\,  \sigma^{\ell_m-2} \widehat{G}_\eta(\sigma) = c R^{\ell_m-1} + \pi \hat{\lambda}_m  \eta^{\half(\Delta_m-\ell_m)} + \cdots
\ee
The dots indicate terms of order $\eta^{\half(\Delta_m - \ell_m)}R$ and $\eta^{\half(\Delta_m-\ell_m)}R^{\ell_m-1}$.  The correction proportional to $\hat{\lambda}_m$ written in \eqref{resec} is meaningful only for $\ell_m>1$, because in this case it dominates over the dots.
Therefore, restricting now to $\ell_m > 1$, we can rearrange the expression $\oint (1 - \widehat{G}_\eta) = 0$ to find
\be\label{sumrule}
\hat{\lambda}_m = \frac{1}{\pi} \lim_{R \to 0} \lim_{\eta \to 0} \eta^{-\half(\Delta_m-\ell_m)} \int_{-R}^R dx  \, x^{\ell_m-2} \mbox{Re}(G_\eta(x) - \widehat{G}_\eta(x)) \ .
\ee
This sum rule is one of our main results. The integrand is related to the commutator $\langle [O,\psi]O\psi\rangle$, and is manifestly positive, as described in section \ref{ss:rbb}.   See \eqref{hatl} for the positive constant relating $\hat{\lambda}_m \propto c_{OOm}c_{\psi\psi m}$.  The role of the first term in the integrand is just to subtract the identity piece, so the integrand can also be written  $x^{\ell_m-2} \mbox{Re\ }(1 - \widehat{G}_\eta(x))$.

\ \\
 \noindent \textit{Comments:}
 
 \subsubsection*{Strict inequality}
 In an interacting theory, we expect $\langle [O,\psi] O \psi\rangle \neq 0$ when $O$ and $\psi$ are timelike separated.  In this case $\lambda_m > 0$ is a strict inequality.

\subsubsection*{Bound on magnitude}
Another way to state the result is using the maximum modulus principle.  (This approach is inspired by the chaos bound, see below.) $\widehat{G}_\eta(\sigma)$ is analytic on $D$, so $|\widehat{G}_\eta(\sigma)|$ cannot have a local maximum.  Combined with the bound \eqref{posrb} on the real line, this implies \eqref{gzze}, with the constant $\sim R^{-\ell_m+1}$. Comparing to \eqref{logex} fixes $\lambda_m>0$ for $\ell_m>1$.

Put differently, if $\lambda<0$, then (considering the stress tensor in $d=4$ for illustration) it is impossible for a function with the expansion
\be
f_\eta(\sigma) = 1 - i \lambda \frac{\eta}{\sigma} + O(\eta^2)
\ee
to be analytic on region $D$ and bounded by 1 on the real line.  An easy way to check this is to plot the magnitude on $\p D$ at radius $R$, then plot the same function on a slightly smaller $\p D$ of radius $R - \delta R$, and compare the maximums.

\subsubsection*{Relation to the chaos bound}
Our derivation is closely related to the bound on chaos recently derived in \cite{Maldacena:2015waa} (see also \cite{Shenker:2013pqa,Roberts:2014isa,Roberts:2014ifa}).  Roughly speaking, our region $D$ plays the role of the complex time strip in \cite{Maldacena:2015waa}. The top and bottom of their strip are analogous (but not equivalent) to the real line segments $\sigma \in [-R,0]$ and $\sigma \in [0,R]$. Where we used the positive expansion coefficients to place various bounds on $|\widehat{G}_\eta(\sigma)|$, they used Cauchy-Schwarz inequalities on thermal correlators.  

The constraints, however, are distinct, and apparently independent. The additional small parameter $\eta \ll 1$ does not appear in the chaos analysis.  This deforms the shape of our contour vs the chaos contour, and is what allows us to place bounds on \textit{perturbative} OPE data.  The chaos bound, on the other hand, bounds the Regge behavior, which in general is not accessible in any OPE.  In holographic CFTs with an Einstein gravity dual, both of these regimes are controlled by the same physics (the bulk graviton).  But in general, the two regimes are controlled by different physics. For example, in a CFT with stringy holographic dual, the causality bound constraints the stress tensor while the chaos bound involves string degrees of freedom.

\subsubsection*{No bound on spin-1 exchange}
We have considered real scalars $O,\psi$, which can have only even-spin  operators in the $t$ channel. The analysis can be generalized to complex scalars and the correlator $\langle \psi O O^\dagger \psi^\dagger\rangle$.  For spin-1 exchange, there is no $1/\sigma$ enhancement of the lightcone block on the second sheet, so the corrections to \eqref{resec} are the same size as the terms that are written and there is no constraint on the spin-1 OPE coefficient.  If there are both spin-1 and spin-2 operators with the same minimal twist, then the spin-2 coefficient still obeys the sum rule and sign constraint.

\subsubsection*{The dominant minimal-twist operator must have spin $\leq 2$}
 Suppose there is a single operator of minimal twist.  Then in region $D$ the $t$ channel expansion for $\eta \ll |\sigma|$ is
\be
\widehat{G}_\eta(\sigma) \sim 1 + i K \eta^{\half(\Delta_m-\ell_m)} \sigma^{1-\ell_m} +\cdots\ ,
\ee
with $K$ a real constant and corrections suppressed by higher powers of $\eta$ or $\sigma$. If $\ell_m>2$, this is simply not possible for a function analytic on $D$, and bounded by 1 on the real line (up to positive powers of $\eta, \sigma$).  This follows easily from the maximum modulus principle: If $\ell_m>2$, then the correction to the magnitude oscillates along the semicircle $S$, so it is impossible to satisfy the requirement
\be
\max_{\p D(R)}|\widehat{G}_\eta(\sigma)| \geq \max_{\p D(R-\delta R)}|\widehat{G}_\eta(\sigma)| \ ,
\ee
where $D(R)$ is the usual region $D$, and $D(R-\delta R)$ is the same region but with a slightly smaller radius. In other words, some directions on the complex $\sigma$ plane fix $K<0$ and others fix $K>0$. 

This is a variation of the chaos bound. The chaos bound constrains the rate of growth of the correlator in the Regge limit, whereas this constrains the growth in the lightcone OPE limit. That these two limits are related in holographic CFTs was already well known, but here we did not assume large $N$. 

It would be very interesting to systematically apply this to large-$N$ CFTs. The stress tensor contribution is $1/N$ suppressed; in some cases this implies other contributions must also be suppressed.  For example, if the minimal-twist operator is the stress tensor and there is an operator of spin $>2$ with next-to-minimal twist, then this operator must also be $1/N$ suppressed.

\subsubsection*{Relation to results of Maldacena and Zhiboedov on Higher Spin Symmetry}

Assuming no scalar operators of dimension $\Delta <d-2$, the stress tensor and other conserved currents are the minimal twist operators.  If there is any finite number of higher spin conserved currents (meaning even $\ell>2$), then we've just shown that the theory violates causality. Therefore, a theory with even one higher spin conserved current must have an infinite number of such currents.  This is a part of the Maldacena-Zhiboedov theorem in $d=3$ \cite{Maldacena:2011jn} and has been demonstrated in higher dimensions using properties of higher spin algebras \cite{Boulanger:2013zza}. Here we have shown it in a way independent of spacetime dimension $d$, and without ever invoking the form of current-current correlation functions.  The assumption that there are no scalars of dimension $\Delta < d-2$ is not actually necessary, as this was shown to hold in a theory with a higher spin current in any $d$ in the first section of  \cite{Maldacena:2011jn}.

\subsubsection*{Stress tensor exchange}
If the minimal twist operator is the stress tensor, then as discussed in section \ref{s:brief} the coefficient $\lambda_m$ is fixed by the conformal Ward identity \cite{Osborn:1993cr}:
\be
\lambda_T \sim \frac{\Delta_O\Delta_\psi}{c} \ ,
\ee
where $c$ is the central charge.  This coefficient (including all the correct prefactors) is obviously positive, so our constraint on the sign does not provide any new information in this case, beyond the satisfaction of linking it to causality and the Regge limit.  

\subsubsection*{Nontrivial sign constraints}
Combining the above comments --- that the sign constraint is absent for $\ell_m < 2$, obvious for $\ell_m = 2$, and $\ell_m > 2$ is excluded --- it might seem that now we cannot derive any interesting sign constraints.  This is wrong; a nontrivial sign constraint for scalars will be discussed in section \ref{s:dphi}. The comments above are evaded in that example because a large parameter suppresses the contribution of the stress tensor relative to other low-twist operators.  We also expect nontrivial inequalities from spinning correlators, as discussed below.

\subsubsection*{Anomalous dimensions}
In \cite{Komargodski:2012ek,Fitzpatrick:2012yx,Fitzpatrick:2014vua,Alday:2014tsa,Vos:2014pqa,Fitzpatrick:2015qma,Alday:2015ota,Kaviraj:2015xsa,Kaviraj:2015cxa,Fitzpatrick:2015zha}, log coefficients in the $t$ channel were related to the anomalous dimensions of high-spin operators exchanged in the $s$ channel. These operators have the schematic form
\be
O_{n,\ell} \sim \psi \Box^n( \p_\mu)^\ell O + O \Box^n( \p_\mu)^\ell \psi +\cdots
\ee
with $\ell \gg 1$.
A positive log coefficient corresponds to a negative anomalous dimension, \ie
\be
\Delta_{n,\ell} < \Delta_O + \Delta_\psi + 2n+\ell \ .
\ee
Not all high-spin operators have negative anomalous dimensions; for example \cite{Komargodski:2012ek}, if $O$ is a charged scalar, then the anomalous dimension of $O \Box^n \p^\ell O$ may have either sign.  This does not contradict our results, since in this case the dominant $t$-channel exchange has spin 1 and our bound does not apply.

\subsubsection*{Light scalars}
The unitarity bound for spinning operators is $\Delta-\ell \geq d-2$. The unitarity bound for scalars allows lower twist, $\Delta \geq \frac{d}{2}-1$. If there is a scalar operator appearing in the $t$ channel that lies in the window
\be\label{window}
\frac{d}{2}-1 < \Delta < d-2 \ ,
\ee
then the lightest such scalar is the minimal-twist operator and dominates in the lightcone limit. The scalar contribution on the second sheet does not grow near the origin, so the coefficient is unconstrained.   It also interferes with our derivation for the spin-2 coefficient, since that is now subleading. Perhaps the scalar contribution can be subtracted to reinstate the spin-2 bound but we leave this to future work.

\subsubsection*{Importance of crossing symmetry}
We used all three crossing channels, $s$, $t$, and $u$ in essential ways.  One way to state the result which makes manifest the connection to permutation symmetry of the Euclidean correlator under swapping $O \leftrightarrow O$ is as follows.  Suppose we are given the spectrum and the $s$ channel OPE coefficients, satisfying positivity, but that this data corresponds to a wrong-sign log term in the $t$ channel.  Then the sum rule will force us to violate positivity in the Regge limit of the $u$ channel, violating unitarity and the $s=u$ crossing relation. This is the crossing relation that comes from swapping $O \leftrightarrow O$ in the Euclidean correlator.

Another way to state the result is in terms of causality.  If a theory is crossing invariant, and the coefficients in the $z$ expansions are positive, but $\lambda_m$ has the wrong sign, then the correlator must have a singularity inside region $D$; this violates both causality and reflection positivity (since it requires the $\rho$ expansion to have negative coefficients).

\subsubsection*{Physical recap}

Despite the technical details, this result is actually very intuitive. Crossing symmetry $s = u$ is the statement that the Euclidean correlator is invariant under swapping the two $O$ insertions.  It is reasonable to expect that this is therefore the key condition preventing the correlator from growing too large, too soon, as $O(x,\bx)$ approaches the lightcone of $O(w,\bw)$.  For the commutator $\langle \Psi |[O, O]|\Psi\rangle$, this only implies that the singularity cannot shift; it does not immediately constrain the finite corrections.  However, even if the singularity does not move, it would be surprising if the subleading finite terms around the singularity grew significantly \textit{larger} as we passed $O(x,\bx)$ through the shockwave generated by $\psi$.  What we have shown is that crossing symmetry, in particular $s=u$, together with reflection positivity do not allow this to happen.

\subsection{Comments on spinning correlators}
Throughout the paper, we have restricted to external scalars, but much of the machinery should carry over to the case of external operators with spin. This will be technically more complicated, but seems likely to lead to interesting bounds on the interactions of spinning fields.  If so, then presumably these are related in some way to the Hofman-Maldacena bounds, derived from an average null energy condition \cite{Hofman:2008ar} (see also a different derivation in \cite{Kulaxizi:2010jt}). 

Consider for example the stress tensor two-point function in a shockwave background, 
\be\label{pttp}
\langle \psi T_{\mu\nu}T_{\rho \sigma} \psi \rangle \ .
\ee
Assuming that the minimal-twist operator appearing in the $TT$ OPE is the stress tensor itself, the log terms in the lightcone limit come with three independent coefficients $n_{1,2,3}$, one for each of the allowed tensor structures in the three-point function $\langle TTT \rangle$.  Our methods will therefore fix the sign of some combination (or combinations) of the $n_i$.  Looking instead at $\langle TTTT\rangle$, we expect to find quadratic (but possibly redundant) constraints on the $n_i$.

The Hofman-Maldacena constraints, in an appropriate basis, are $n_i \geq 0$. In CFTs with a holographic dual, the same constraints were derived from causality on the gravity side, in the background of a gravitational shockwave \cite{Brigante:2008gz,Hofman:2009ug,Camanho:2009vw}. In a general CFT, the connection between causality and the Hofman-Maldacena constraints remains unclear.  Perhaps our methods can be extended to address this puzzle.

\section{Holographic dual of the $(\p \phi)^4$ constraint}\label{s:dphi}
In the theory of a massless scalar with a shift symmetry,
\be\label{scalars}
S = \int d^Dx  \sqrt{-g}\left[-(\del \phi)^2 + \mu (\del^\mu \phi \del_\mu \phi)^2  + \cdots \right]\ ,
\ee
causality enforces \cite{Adams:2006sv}
\be\label{mubound}
\mu \geq 0 \ .
\ee
The coupling is not renormalizable, so this is an effective theory; if $\mu < 0$, it cannot be UV-completed. This constraint also plays an essential role in the proof of the $a$ theorem for renormalization group flows in four dimensions. In that context, $S$ is the effective action of the dilaton, and $\mu = a_{UV} - a_{IR}$ \cite{Komargodski:2011vj}.

In this section, we view \eqref{scalars} as a theory in anti-de Sitter space, and show that our constraint can be used to derive \eqref{mubound} from the dual CFT. In this section --- and only in this section --- we assume large $N$.  The result of \cite{Adams:2006sv} was in flat space, whereas we will derive the constraint for the same Lagrangian in AdS.  It holds for any value of the AdS radius, so this suggests the flat-space bound as well, but we will not address whether we can strictly set $R_{AdS} =\infty$.\footnote{We should also remark that there is some debate over whether the results of \cite{Adams:2006sv} indicate that a theory with $\mu<0$ is unacceptable, or that the theory is acceptable but the causality-violating solutions are unphysical \cite{Bruneton:2006gf}. See \cite{Papallo:2015rna} for an elucidating discussion, of both scalars and Gauss-Bonnet gravity.  The basic point (in the non-gravitational context) is that a theory can be causal but superluminal. In any case, we will show that a scalar theory in AdS with the wrong sign cannot be embedded into a UV theory that is dual to a CFT obeying the usual Euclidean axioms.
}

In our approach the action \eqref{scalars} is the bulk theory, but it does not include gravity. In the dual CFT, this means we are taking the central charge $c \to \infty$ to decouple the stress tensor, while holding fixed the large parameter $N$ that suppresses the connected correlators of other CFT operators.  This is not consistent in the UV, but defines a reasonable effective theory in the IR, both in the bulk and on the boundary. (See \cite{Fitzpatrick:2010zm} for the boundary point of view.) The holographic relationship between \eqref{scalars} and CFT is not full blown holography, which is a statement about quantum gravity that makes sense only at finite (but small) Newton's constant, but it is a nontrivial subsector.

We cannot directly apply the sum rule \eqref{sumrule} because there we assumed a single operator of minimal twist, but the bound on the log coefficient can be applied with minor modifications. The basic idea is simple.\footnote{We thank A. Zhiboedov for a discussion that led to this section.}  At $\mu=0$, the bulk theory \eqref{scalars} is dual to a generalized-free CFT with an operator of integer dimension $d$. The 4-point function is analytic, so all of the logs in the conformal block expansion must cancel after summing over primaries.  Now turning on the $(\p \phi)^4$ interaction will introduce a log, with coefficient proportional to $\mu$, from the exchange of the spin-2 operator $O \p_\mu \p_\nu O$.  Then our bound fixes the sign of $\mu$.  The main thing we need to check is that the result comes out with the right sign.

\subsection{Bootstrap}
Following \cite{Heemskerk:2009pn}, we need to map the bulk interaction into an effect on CFT data.
At $\mu = 0$, a free scalar in the bulk is dual to a generalized free theory on the boundary.  Connected correlators vanish, so the four-point function is the sum over channels of the identity contribution:
\be\label{ltff}
G(z,\bz) \equiv \langle O(0) O(z,\bz) O(1) O(\infty) \rangle = 1 + (z\bz)^{-\Delta_O} + [(1-z)(1-\bz)]^{-\Delta_O} + O(\mu) \ ,
\ee
where $O$ is the operator dual to $\psi$. The dimension of an operator dual to a massless scalar is
\be
\Delta_O = d \ .
\ee
\eqref{ltff} obviously solves the crossing equation. Before turning on the perturbation, let's express \eqref{ltff} as an expanion in the $s$ channel.  At $\mu = 0$, 
\be\label{scalarsum}
G(z,\bz) = (z \bz)^{-d}\left[1 + \sum_{n \geq 0,\ell \geq 0}c_0(n,\ell)^2 g_{\Delta_0(n,\ell), \ell}(z,\bz) \right]\ .
\ee
The sum is over double-trace operators, schematically $O_{n,\ell} \sim O \Box^n (\p_\mu)^\ell O$, with dimension
\be
\Delta_0(n,\ell) = 2d + 2n  + \ell \ .
\ee
The OPE coefficients $c_0(n,\ell)$ are known but will not be needed, beyond the fact that they are real.

Now,  still following \cite{Heemskerk:2009pn} and working to $O(\mu)$, the bulk coupling $\mu(\del \phi)^4$ leads to a slightly deformed solution of crossing, with small corrections to both $c_0(n,\ell)$ and $\Delta_0(n,\ell)$.  Writing the perturbed dimensions as
\be
\Delta(n,\ell) = \Delta_0(n,\ell)+ \gamma(n,\ell) \ ,
\ee
the perturbed OPE coefficients are $c(n,\ell) = c_0(n,\ell) + \delta c(n,\ell)$ with \cite{Heemskerk:2009pn,Fitzpatrick:2011dm}
\be
c_0(n,\ell)\delta c(n,\ell) = \frac{1}{4} \frac{\p}{\p n}\left[ c_0(n,\ell)^2 \gamma(n,\ell)\right] \ .
\ee
The $(\del \phi)^4$ coupling turns on anomalous dimensions $\gamma(n,\ell)$ only for $\ell=0,2$ \cite{Heemskerk:2009pn}.

\subsection{Logs}

Now we want to apply the bound derived in section \ref{s:logs}. Since $\Delta_O=d$ is an integer, the unperturbed four-point function is analytic; there are no logs.  On the other hand, individual contributions to \eqref{scalarsum} in the limit $\bz \to 1$ have the usual logs coming from the lightcone conformal block.  This is consistent only if all of the logs cancel in the $\mu=0$ sum, so we can drop the $O(\mu^0)$ term and focus on the perturbation.

The leading correction in the lightcone limit comes from operators of minimal twist, \ie $n=0$. Only the scalar and spin-2 contributions $\gamma(n,0)$ and $\gamma(n,2)$ are non-zero. The dominant terms near the lightcone on the second sheet come from the highest spin, so the only contribution we need to consider is from the operator with $n=0$, $\ell=2$:
\be
O_{0,2} \sim O \p_\mu \p_\nu O \ .
\ee
Defining the normalized correlator
\be
\overline{G}(z,\bz) = [(1-z)(1-\bz)]^{d}G(z,\bz) \ ,
\ee
the order-$\mu$ contribution to $\overline{G}(z,\bz)$ from this operator in the $t$ channel is
\be
\delta \overline{G}(z,\bz) = \frac{1}{2}\p_n\left[ c_0(n,\ell)^2 \gamma(n,\ell) g_{2\Delta_O+2n+2, 2}(1-z,1-\bz)\right]_{n=0} \ .
\ee
The largest contribution as $\bz \to 1$ comes from the derivative acting on $(1-\bz)^{\Delta_O + n}$,
\be
\delta \overline{G} \sim \frac{1}{2}c_0(0,2)^2\gamma(0,2)(1-\bz)^{d}\log(1-\bz)\tilde{g}_{2d+2,2}(1-z)
\ee
where $\tilde{g}$ is the lightcone block \eqref{lcblockf}. On the first sheet, the leading term as $z \to 1$ is
\be
\delta \overline{G} \sim 2 c_0(0,2)^2\gamma(0,2) \log(1-\bz) (1-\bz)^d(1-z)^{d+2} \ .
\ee
As expected, this correction is highly suppressed in the shockwave kinematics, $\delta \overline{G} =O( \delta^{2d+2}\log\delta)$. 
On the second sheet, defined by sending $z \to e^{-2\pi i }z$ within the lightcone block, the leading term as $z \to 1$ is
\be
\delta \overline{G} \sim -\frac{ i \pi}{4} c_0(0,2)^2\gamma(0,2) (1-\bz)^d \log(1-\bz) (1-z)^{-d-1} \frac{\Gamma(2d+4)^2}{(2d+3)\Gamma(d+2)^4} \ .
\ee
In the shockwave kinematics this terms grows as $\delta^{-1}$, so the sign of the coefficient is fixed by the arguments of section \ref{s:logs}. To find the correct sign, we set $z = 1 + i\ep$, $\bz = 1 + i \bvep$, and expand for real $\ep,\bvep$ with $0 \ll \bvep \ll \ep \ll 1$:
\be\label{ffog}
\delta \overline{G} \sim c_0(0,2)^2 \gamma(0,2) \frac{\bvep^d}{\ep^{d+1}} \log \left(\frac{1}{\bvep}\right) \times (\mbox{positive}) \ .
\ee
The bound derived in section \ref{s:logs}, equation \eqref{gzze}, is
\be
|1 + \delta \overline{G}| \leq 1 
\ee
(up to terms suppressed by positive powers of $\ep, \bvep$). Comparing to \eqref{ffog}, this implies that the anomalous dimension must be negative
\be\label{gams}
\gamma(0,2) \leq 0 \ .
\ee
This CFT bound agrees with the bulk bound $\mu \geq 0$.  To check the sign in a convention-independent way, we compare to  \cite{Alday:2014tsa}, where the anomalous dimensions in the perturbative solution of crossing are related to the flat-space $S$-matrix of supergravity. To make the comparison, we can set the supergravity contributions to zero (in their eqn (86)). Then an $S$-matrix that behaves in the forward limit $t \to 0$ as $\mathcal{M}(s,t) = \alpha s^2 + O(s^4)$ is dual to $\gamma(0,2) = -\frac{10}{11}\alpha$. The bulk constraint of \cite{Adams:2006sv} is $\alpha>0$, so \eqref{gams} has the correct sign.

\para{Acknowledgments}
It is a pleasure to thank Sumit Das, Liam Fitzpatrick, Daniel Harlow, Jared Kaplan, Juan Maldacena, Gautam Mandal, Shiraz Minwalla, Alex Maloney, David Poland, Dan Roberts, Steve Shenker, David Simmons-Duffin, Douglas Stanford, John Stout, and Alexander Zhiboedov for helpful discussions; Shiraz Minwalla for comments on a draft; and Alexander Zhiboedov for comments on a draft as well as helpful suggestions incorporated in $v2$. This work is supported by DOE Early Career Award DE-SC0014123. The work of SK is supported by NSF grant PHY-1316222.

\appendix

\section{Positive coefficients in the $s$ and $u$ channels}\label{app:positive}
This appendix fills in the details of section \ref{ss:positivity}, showing that the coefficients of $z^h \bz^{\bh}$ in the $s$ channel expansion are positive.

In \cite{Fitzpatrick:2012yx} it was shown that individual conformal blocks with equal external weights have an expansion with positive coefficients (up to an overall sign, which in our conventions is $(-1)^{\ell_p}$).  We will apply a similar argument to the full correlator, allowing for $\Delta_O \neq \Delta_\psi$. This automatically accounts for both the blocks and the OPE coefficients.   
Define
\be\label{fstateapp} 
|f\rangle = \int_{\epsilon}^{1-\epsilon} dr_1 \int_0^{2\pi}d\theta_1 r_1^{\Delta_O+\Delta_\psi} f(r_1,\theta_1)O(r_1e^{i \theta_1}, r_1 e^{-i\theta_1})\psi(0)|0\rangle \ ,
\ee
where $\epsilon>0$ is small, and $f$ is any function that is smooth on the domain of integration. The conjugate, defined by inversion across the unit sphere, is
\be
\langle f| = \langle 0| \psi(\infty)  \int_\epsilon^{1-\epsilon} dr_2 \int_0^{2\pi}d\theta_2 \,  r_2^{-\Delta_O+\Delta_\psi} f^*(r_2,\theta_2)O(\frac{1}{r_2}e^{i\theta_2}, \frac{1}{r_2}e^{-i\theta_2}) \ .
\ee
Reflection positivity requires states to have positive norms, so 
\begin{align}\label{firstnorm}
\int_\epsilon^{1-\epsilon}dr_1 \int_\epsilon^{1-\epsilon} dr_2 \int_0^{2\pi}&d\theta_1 \int_0^{2\pi}d\theta_2  \, r_2^{-2\Delta_O} (r_1 r_2)^{\Delta_O+\Delta_\psi}  \times \\
&f(r_1,\theta_1)f^*(r_2,\theta_2) \langle \psi(0) O(x,x^*) O(y,y^*) \psi(\infty)\rangle > 0 \notag
\end{align}
where $x = r_1 e^{i \theta_1}, y=\frac{1}{r_2}e^{i\theta_2}$. Now, by a conformal transformation, the four-point function in the integrand can be related to the canonical insertion points, 
\bea
\langle \psi(0) O(x,x^*) O(y,y^*) \psi(\infty)\rangle &=& r_2^{2\Delta_O} G(\frac{x}{y},\frac{x^*}{y^*}) \\
&=& \frac{r_2^{\Delta_O-\Delta_\psi}}{ r_1^{\Delta_O+\Delta_\psi}} \sum_{h,\bh} a_{h,\bh} (r_1r_2)^{h+\bh}e^{i(h-\bh)(\theta_1-\theta_2)} \ .\notag
\eea
Therefore, absorbing the regulator $\epsilon$ into the definition of $f$, \eqref{firstnorm} becomes
\be
\sum_{h,\bh \geq 0}a_{h,\bh} \left| \int_0^1 dr \int_{0}^{2\pi} d\theta \, r^{h+\bh} e^{i(h-\bh)\theta}f(r,\theta)\right|^2 > 0 
\ee
for any function $f$ such that this converges. Choosing $f(r,\theta) = r^{-m-1} e^{i m \theta}f(r)$ projects onto a single spin, $\bh = h + m$, leaving
\be
\sum_{h\geq 0} a_{h, \bh} \left| \int_0^1 dr \, r^{2h-1}f(r) \right|^2 > 0 \ .
\ee
Writing $r=e^{-\lambda}$, the integral is a Laplace transform. We can choose $f(r)$ to project onto a particular value of $h$.  For example, choose $f$ such that $\int_0^1 dr r^{2h-1} f(r) = \left( 2h h_0\over h^2+h_0^2\right)^N$. This is possible for any integer $N$ by an inverse Laplace transform, and for $N$ large it is very strongly peaked at $h=h_0$. Thus $a_{h,\bh}>0$.

A check is to note that individual blocks in $d=4$ have a positive expansion in $z,\bz$, up to a possible overall factor of $(-1)^{\ell_p}$. This implies that either $\langle \psi(0) O(z,\bz)O(1) \psi(\infty) \rangle$ or $\langle\psi(0) O(z,\bz) \psi(1) O(\infty)\rangle$ has a positive expansion, since these differ only by $c_{O\psi p} \to c_{\psi O p} = c_{O \psi p}(-1)^{\ell_p}$ in front of each block.  To decide which of these orderings has a positive expansion, we can just check one example: two decoupled scalars with equal dimension $\Delta = \Delta_O = \Delta_\psi$.  The four-point function expanded in the $s$ channel is\bea
\langle \psi(x_1) O (x_2) O(x_3) \psi(x_4)\rangle &=& (x_{14}^2 x_{23}^2)^{-\Delta}\\
&=& (x_{12}^2 x_{34}^2)^{-\Delta} \left( z \bz\over (1-z)(1-\bz)\right)^{\Delta}\notag
\eea
This has positive coefficients in the $z$ expansion. In $d=4$, the decomposition into conformal blocks (see appendix \ref{app:doblocks}) also has the expected signs, 
\be
 \left( z \bz\over (1-z)(1-\bz)\right)^{\Delta} = g_{2\Delta, 0}^{0,0}(z,\bz) - \frac{\Delta}{2}g_{2\Delta+1,1}^{0,0} (z,\bz)+ \cdots \ .
 \ee
 
 \section{Conformal blocks in $d=4$}\label{app:doblocks}

 Our convention for the full conformal block for external scalars in $d=4$ is \cite{Dolan:2000ut}:
 \be
 g_{\Delta, \ell}^{\Delta_{12}, \Delta_{34}}(z,\bz) = (-2)^{-\ell}\frac{z^h \bz^{\bh+1}}{\bz-z} F(h-1, z)F(\bh, \bz) + (z \leftrightarrow \bz)
 \ee
 with
 \be
 h = \half(\Delta - \ell) , \qquad \bh = \half(\Delta+\ell) \ ,
 \ee
 and
 \be
 F(h, z) \equiv \, _2F_1(h-\half \Delta_{12}, h + \half \Delta_{34}, 2h, z) \ .
 \ee

\end{spacing}

\end{document}